\documentclass[fleqn,10pt]{wlscirep}
\usepackage[utf8]{inputenc}
\usepackage[T1]{fontenc}

\usepackage[capitalise]{cleveref}
\usepackage{subcaption} 
\usepackage{bm} 
\usepackage{mathtools}  
\usepackage{amsmath,amsfonts}
\usepackage{tikz}
\usepackage{nicefrac}   
\usepackage[normalem]{ulem}

\DeclareMathAlphabet\mathbfcal{OMS}{cmsy}{b}{n} 
\usetikzlibrary{}
\newcommand{\fancyR}{\sbox1{\vbox{R}}\sbox2{\hbox{R}}\tikz[inner sep=0pt,outer sep=0pt]{\coordinate (A);\draw[-,black,line width=0.55pt,scale=0.75]([shift={({\the\wd2/2},0)}]A) to[out=180,in=0] ++(-{\the\wd2/2},{3*(\the\ht1+\the\dp1)/5)}) to[in=90,out=180]++({-\the\wd2/5},{-(\the\ht1+\the\dp1)/8})
to[in=270,out=270]++({\the\wd2/2},{7*(\the\ht1+\the\dp1)/12})
to[in=0,out=90]++(-{7*\the\wd2/20},{3*(\the\ht1+\the\dp1)/12})
to[in=90,out=180]++(-{13*\the\wd2/24},-{11*(\the\ht1+\the\dp1)/12})
to[in=180,out=270]++({3*\the\wd2/12},{-4*(\the\ht1+\the\dp1)/10})
to[in=270,out=0]++({11*\the\wd2/48},{(\the\ht1+\the\dp1)/3})
to[in=300,out=90]++(-{3*\the\wd2/13},{11*(\the\ht1+\the\dp1)/12})
to[in=40,out=120]++(-{6*\the\wd2/10},-{1*(\the\ht1+\the\dp1)/6});
}}

\title{KIDS: kinematics-based (in)activity detection and segmentation in a sleep case study}

\author[1]{Omar Elnaggar}
\author[2]{Roselina Arelhi}
\author[3]{Frans Coenen}
\author[4]{Andrew Hopkinson}
\author[5,6]{Lyndon Mason}
\author[1,*]{Paolo Paoletti}
\affil[1]{School of Engineering, University of Liverpool, Liverpool L69 3GH, United Kingdom}
\affil[2]{Faculty of Engineering, University of Sheffield, Sheffield S1 3JD, United Kingdom}
\affil[3]{School of Electrical Engineering, Electronics and Computer Science, University of Liverpool, Liverpool L69 3BX, United Kingdom}
\affil[4]{School of Psychology,University of Liverpool, Liverpool L69 7ZA, United Kingdom}
\affil[5]{School of Medicine, University of Liverpool, Liverpool L69 3GE, United Kingdom}
\affil[6]{Department of Trauma and Orthopaedics, Liverpool University Hospitals NHS Foundation Trust, Liverpool L9 7AL, United Kingdom}

\affil[*]{P.Paoletti@liverpool.ac.uk}

\begin{abstract}
Sleep behaviour and in-bed movements contain rich information on the neurophysiological health of people, and have a direct link to the general well-being and quality of life. Standard clinical practices rely on polysomnography for sleep assessment; however, it is intrusive, performed in unfamiliar environments and requires trained personnel. Progress has been made on less invasive sensor technologies, such as actigraphy, but clinical validation raises concerns over their reliability and precision. Additionally, the field lacks a widely acceptable algorithm, with proposed approaches ranging from raw signal or feature thresholding to data-hungry classification models, many of which are unfamiliar to medical staff. This paper proposes an online Bayesian probabilistic framework for objective (in)activity detection and segmentation based on clinically meaningful joint kinematics, measured by a custom-made wearable sensor. Intuitive three-dimensional visualisations of kinematic timeseries were accomplished through dimension reduction based preprocessing, offering out-of-the-box framework explainability potentially useful for clinical monitoring and diagnosis. The proposed framework attained up to 99.2\% $F_1$-score and 0.96 Pearson’s correlation coefficient in, respectively, the posture change detection and inactivity segmentation tasks. The work paves the way for a reliable home-based analysis of movements during sleep which would serve patient-centred longitudinal care plans.
\end{abstract}
\begin{document}

\flushbottom
\maketitle
%
%
\thispagestyle{empty}


\section*{Introduction}

The study of a human sleep behaviour reveals their state of health and well-being. Habitual in-bed behaviour can reveal physiological and neurological disorders that are otherwise latent during wakefulness \cite{Ibanez2018} such as {\itshape restless leg syndrome} and {\itshape periodic leg movements}.
{\color{black} Sleep deprivation and intermittent sleep were found to be linked to multiple health risks \cite{Beccuti2011,Calhoun2010,Spiegel2005,Wolk2005}.}
In-bed sleep behaviour (movements and postures) can sometimes also cause health complications{\color{black}, such as {\itshape pressure sores} \cite{Paquay2008}, apnoea \cite{Pinna2015} and painful spasms \cite{Akeson1987,Parisi2003}.}

In the light of the clinical context outlined above, there has been a growing interest within the research community to study human sleep behaviour. Different aspects have been investigated including sleep posture classification \cite{Elnaggar2020,Elnaggar2022a}, detection of in-bed movements and posture transitions \cite{Alaziz2020,Jeon2019}, sleep staging\cite{Perslev2021}, sleep physiology and vital sign monitoring{\color{black}\cite{Yang2017,Ben-Dov2007}.}
Various technologies have been employed for at-home and in-clinic sleep monitoring. The clinical gold standard for the assessment of sleep-related disorders is {\itshape polysomnography} (PSG) which measures multiple {\color{black} physiological} parameters.
{\color{black} There are, however, disadvantages to using PSG such as} sensor and electrode intrusiveness, unfamiliar sleep environment, and cost of personnel training and technology. Therefore,  alternatives to PSG have been proposed to make less sophisticated sleep assessments.
Popular options included the less intrusive {\itshape accelerometer-based sensing} (actigraphy \cite{Chang2018}) which involves an actigraphic device, such as a smartwatch worn around the wrist or ankle, to record motor activity during sleep and measure parameters like sleep quality and duration. Other solutions adopted {\itshape bed-embodied sensors}, such as load cells \cite{Alaziz2020}, and {\itshape in-bedroom sensors} such as app-empowered smartphones \cite{Min2014} which incorporates multiple sensors like accelerometers and microphones.


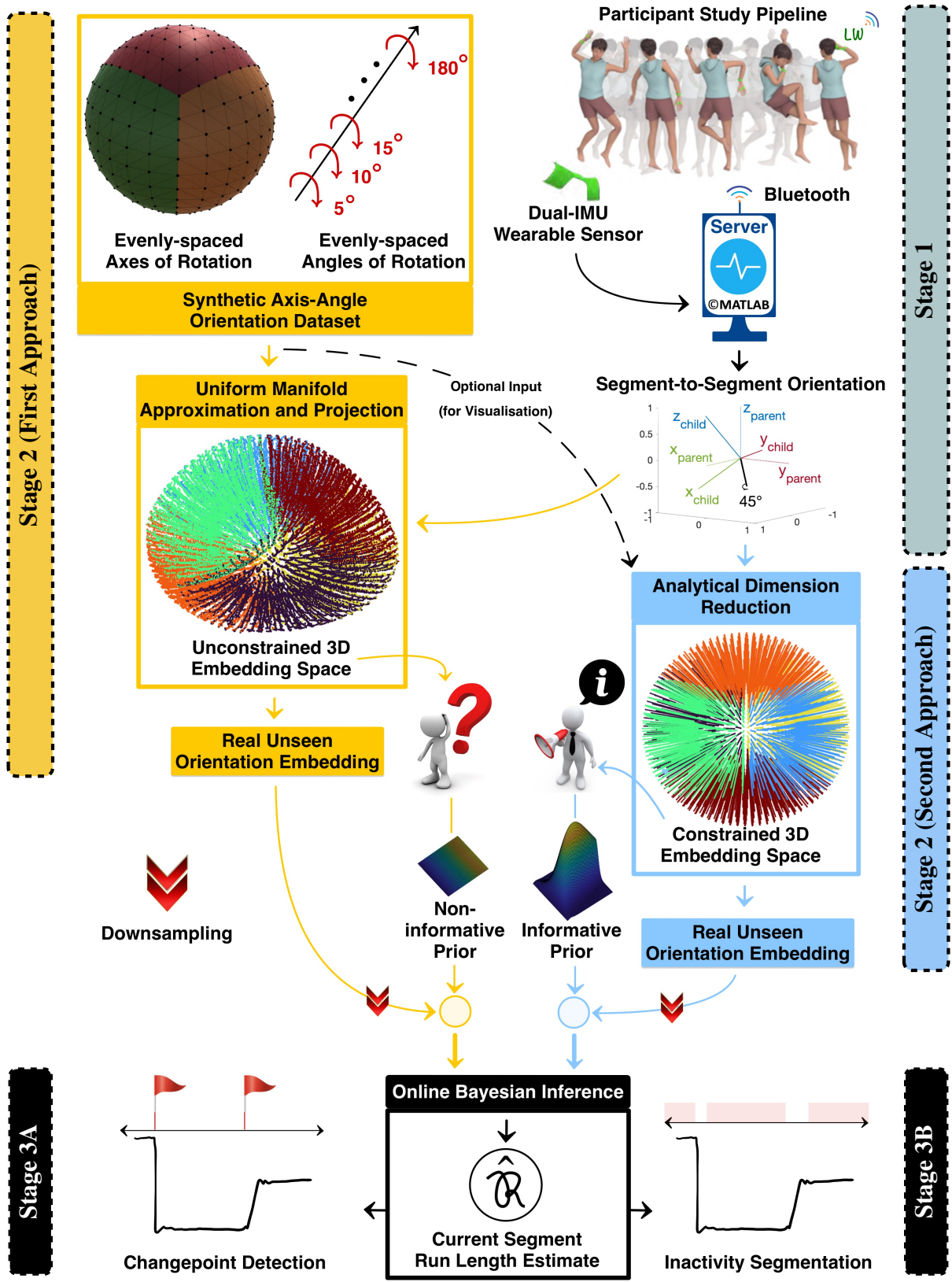
\begin{figure}[!t]
    \centering
\begin{tikzpicture}
\node[above right] (MainFigure) at (0,0) {\includegraphics[height=0.92\textheight]{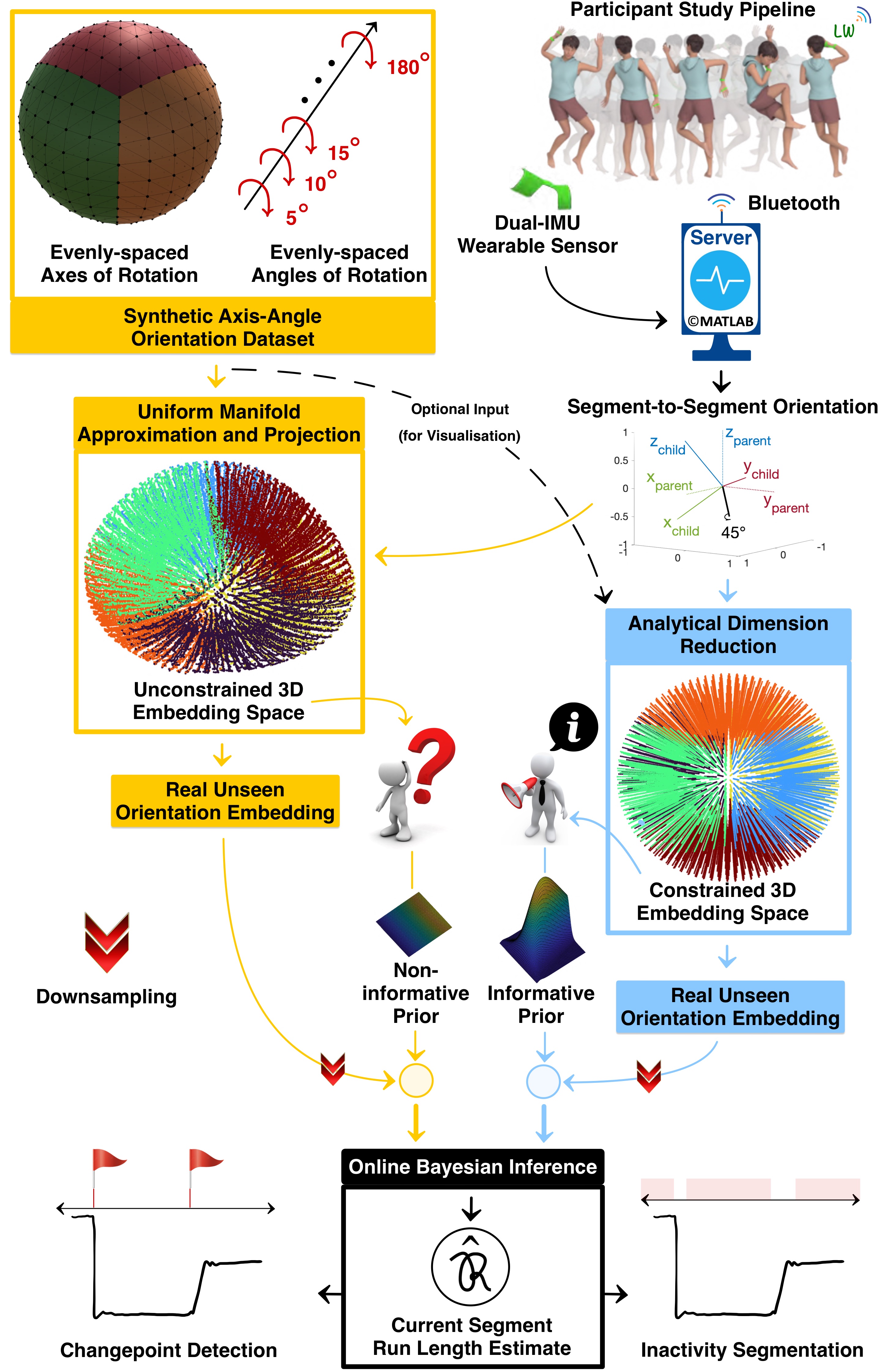}};

\definecolor{tempcolor1}{RGB}{245,202,69}
\definecolor{tempcolor2}{RGB}{177,200,202}
\definecolor{tempcolor3}{RGB}{151,199,249}
\definecolor{tempcolor4}{RGB}{0,0,0}

\draw[ultra thick,rounded corners,black,dashed,fill=tempcolor2] (0.81\linewidth,0.53\textheight) rectangle (0.81\linewidth+0.75cm,0.925\textheight) node[pos=.5,align=center,rotate=90] {\hypertarget{fig:kids_framework_a}{\large \bfseries Stage 1}};

\draw[ultra thick,rounded corners,black,dashed,fill=tempcolor1] (-0.05\linewidth,0.37\textheight) rectangle (-0.05\linewidth+0.75cm,0.92\textheight) node[pos=.5,align=center,rotate=90] {\hypertarget{fig:kids_framework_b}{\large \bfseries Stage 2 (First Approach)}};

\draw[ultra thick,rounded corners,black,dashed,fill=tempcolor3] (0.81\linewidth,0.23\textheight) rectangle (0.81\linewidth+0.75cm,0.52\textheight) node[pos=.5,align=center,rotate=90] {\hypertarget{fig:kids_framework_c}{\large \bfseries Stage 2 (Second Approach)}};

\draw[ultra thick,rounded corners,white,dashed,fill=tempcolor4] (-0.05\linewidth,0.01\textheight) rectangle (-0.05\linewidth+0.75cm,0.16\textheight) node[pos=.5,align=center,rotate=90] {\hypertarget{fig:kids_framework_d}{\large \bfseries \color{white}{Stage 3A}}};

\draw[ultra thick,rounded corners,white,dashed,fill=tempcolor4] (0.81\linewidth,0.01\textheight) rectangle (0.81\linewidth+0.75cm,0.16\textheight) node[pos=.5,align=center,rotate=90] {\hypertarget{fig:kids_framework_d}{\large \bfseries \color{white}{Stage 3B}}};

\end{tikzpicture}

\caption{Graphical illustration of the proposed kinematics-based (in)activity detection and segmentation (KIDS) framework.}
\label{fig:kids_framework}
\end{figure}

{
\color{black}
Within the large field of in-bed movement analysis, there are commonly three research directions reported in the literature: {\itshape active/idle state detection}, {\itshape wake/sleep state detection} and {\itshape sleep stage estimation}. From the literature, these directions broadly rely on similar methodologies, namely threshold-based, classification-based and hybrid approaches. Threshold-based approaches \cite{Alaziz2016,Alaziz2020,Gu2014,Gu2016a,Borazio2014,Chang2018,Kwasnicki2018,Nakazaki2014,Sun2017} are the most popular and they apply a predefined threshold hyperparameter to a predictor variable (raw data or processed features) to segment sensor timeseries. Classification-based approaches \cite{Min2014,Gu2014,Gu2016a,Banfi2021} employ classifiers to identify which states, e.g. active or idle, sensor measurements belong to. The less popular hybrid approaches \cite{Chang2018,Domingues2014} use a mixture of threshold- and classification-based approaches; a thresholding algorithm typically produces preliminary labels which are then refined by a classifier to improve performance.
The previous approaches have shortcomings, such as the detection of short-lasting wake state surrounded by long-lasting sleep state \cite{Palotti2019}. Therefore, it is common in the literature to employ handcrafted ``re-scoring rules'' to correct for such systematic errors \cite{Webster1982}. Nevertheless, this set of rules need to be applied with caution as they may favour accuracy over $F_1$ score or even degrade both of them \cite{Palotti2019}. Additional insights can be found in comparative studies \cite{Olivares2012,Palotti2019} which analysed the performance of previous approaches with focus on actigraphy.
}

The limitations of existing work on body movement analysis during sleep can therefore be summarised as follows. On one hand, threshold-based approaches have tricky-to-tweak hyperparameter(s), as different subjects exhibit varying in-bed behaviour and movement intensities. These approaches also rely on raw sensor data or manually extracted features which are not necessarily the best representation of information for movement analysis. On the other hand, classification-based approaches require large-size datasets for classifier training, and these datasets are typically imbalanced in nature (disproportionate class-wise sample size). Another issue is the performance dependency of classifiers on the dataset population which lack diversity of ethnicity, age, etc \cite{Palotti2019}.


This paper presents a novel {\itshape kinematics-based (in)activity detection and segmentation} (KIDS) framework (depicted in \cref{fig:kids_framework}) based on {\itshape dimension reduction} (DR) and Bayesian inference, {\color{black}without the need for na\"{i}ve hard rules or longitudinal collection of unbalanced training data.}
KIDS leverages {\itshape Bayesian inference} to: (i) perform probabilistic modelling of preprocessed joint kinematic timeseries, (ii) objectively detect the temporal locations of posture changepoint events and (iii) segment preprocessed timeseries into segments of inactivity according to the estimated Bayesian statistics, i.e. mean and variance. The decision making of the KIDS framework is primarily based on the current segment run length, $\fancyR$, probabilistically determined at each time instant. Inactivity segments contain kinematic observations of consistent statistics which allow $\fancyR$ to (linearly) increase in value, whereas activity periods typically come with abrupt and permanent change in the estimated statistics that result in repetitive resetting of $\fancyR$ to zero. On the user end, a semi-automatic {\itshape reset detection logic} with an adjustable duration threshold parameter is available to detect changepoints and segment the timeseries data into sleep postures according to specific needs, without altering the underlying Bayesian inference of (in)activity. This paper demonstrates a possible use of this parameter to select sufficiently long segments while ruling out short-lasting ones associated with periods of activity. However, clinical users may opt for different uses as per their needs, for example, to extract short-lasting segments instead. The disentanglement of the threshold parameter from the Bayesian perception of (in)activity highlights a fundamental difference from existing threshold-based approaches where thresholding is the backbone of (in)activity perception.  

KIDS is a generalised framework that exploits joint kinematics. In this work, the {\itshape left wrist} (LW) joint is considered for implementation, but other body joints could be substituted. The LW joint kinematics describe the hand-to-forearm orientation and were captured by a custom-made miniature wearable sensor module with two embedded {\itshape inertial measurement units} (IMUs).
The measured orientations is natively represented in the {\itshape four-dimensional} (4D) {\itshape axis-angle space}, which is not readily visualisable and likely unfamiliar to non-technical medical experts. Therefore, two DR methods were employed to preprocess the axis-angle space, mapping the wrist kinematics to a three-dimensional (3D) space, ready for direct visualisation and subsequent Bayesian inference at lower computational complexity. The first dimension reduction method produced an unconstrained 3D embedding space using {\itshape Uniform Manifold Approximation and Projection} (UMAP) \cite{McInnes2018}. The second method was a proposed analytical approximation of the UMAP embedding space which produces a fully constrained 3D embedding space. We compare both 3D embedding spaces, and discuss the implications these have on the Bayesian inference and the overall performance of the (in)activity detection and segmentation.

The choice of segment-to-segment kinematics was motivated by the authors' recent work which showed the effectiveness of using similar kinematic cues from four extremity joints (wrists and ankles) in recognising twelve sleep postures \cite{Elnaggar2022a}. This previous work required manual segmentation of each posture to showcase the posture classification performance. The KIDS framework proposed in this paper provides autonomous (in)activity detection and segmentation, which have great potential to empower several clinical applications, including in-bed posture analysis and sleep behaviour disorder screening. Even though kinematic measurements from the wrists and ankles were necessary to discriminate between postures, this paper shows evidence that the kinematic profile of the LW alone is sufficient for the (in)activity detection and segmentation task.

\section*{Results}

The proposed KIDS framework, shown in \cref{fig:kids_framework}, involves three stages: (i) wearable inertial sensing for the measurement of segment-to-segment orientation across the LW joint, (ii) DR-based joint kinematics preprocessing and visualisation, and (iii) kinematics-based Bayesian inference for (in)activity detection and segmentation. Presented in this section are highlights from each stage.

\subsection*{Simulated Sleep Protocol and Wrist Kinematics Measurement}

\begin{figure}[!t]
    \centering
    \includegraphics[width=\linewidth]{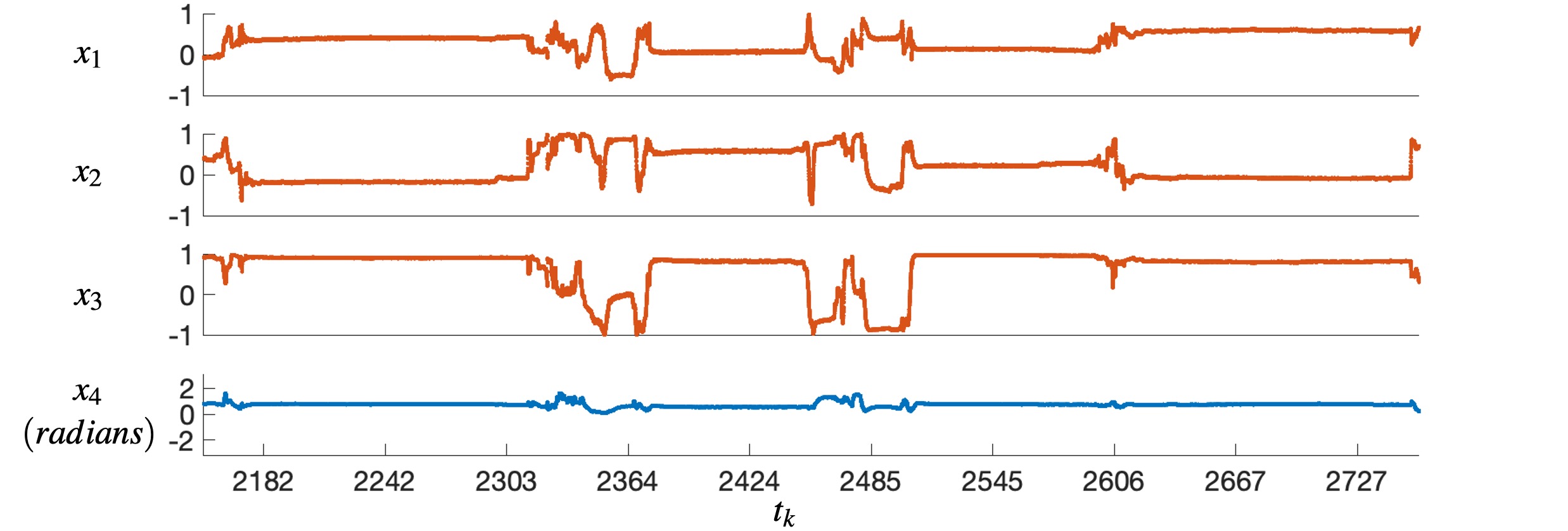}
    \caption{A timeseries describing the hand-to-forearm orientation in the axis-angle space.}
    \label{fig:LW_axisangle_timeseries}
\end{figure}

A simulated sleep experimental protocol {\color{black}(discussed in the Methods Section)} was devised to validate the proposed KIDS framework.
{\color{black} The protocol emulates real sleep by guiding participants through a sequential replication of twelve common sleep postures in a shuffled order.
The collected inertial measurements from wearable sensors were subsequently fused to estimate the hand-to-forearm orientation in the form of a quaternion, which was subsequently converted to the 4D axis-angle representation (shown in \cref{fig:LW_axisangle_timeseries}) for subsequent preprocessing and joint kinematics visualisation.}
In this paper, ${\bm x} \in \mathbb{R}^4$ $(= x_1 \cdot \hat{i} + x_2 \cdot \hat{j} + x_3 \cdot \hat{k} + x_4 \cdot \hat{w})$ represents the sensor-measured segment-to-segment orientation in the axis$(\hat{i},\ \hat{j},\ \hat{k})$-angle$(\hat{w})$  space. The relative orientation timeseries is indexed using a timestamp vector ${\bm t} = t_1, t_2, ..., t_T$.

\subsection*{Joint Kinematics Preprocessing and Visualisation}


Enabled by two different DR methods, UMAP-based and ADR-based, the preprocessing stage produces a reduced dimensional representation of sensor-measured joint orientation, ${\bm x}$, allowing for intuitive 3D visualisation potentially beneficial to non-technical medical experts, while lowering the computational complexity of subsequent Bayesian inference. The output preprocessed orientation embedding is mathematically denoted by ${\bm o} \in \mathbb{R}^3$ $(= o_1 \cdot \hat{i} + o_2 \cdot \hat{j} + o_3 \cdot \hat{k})$. The complete embeddings dataset $\mathbfcal{O}$ encloses all preprocessed ${\bm o}$ over the timestamp vector, ${\bm t}$.

\begin{figure}[!t]
    \centering
    \begin{subfigure}[b]{0.49\textwidth}
        \centering
    	\includegraphics[width=\linewidth]{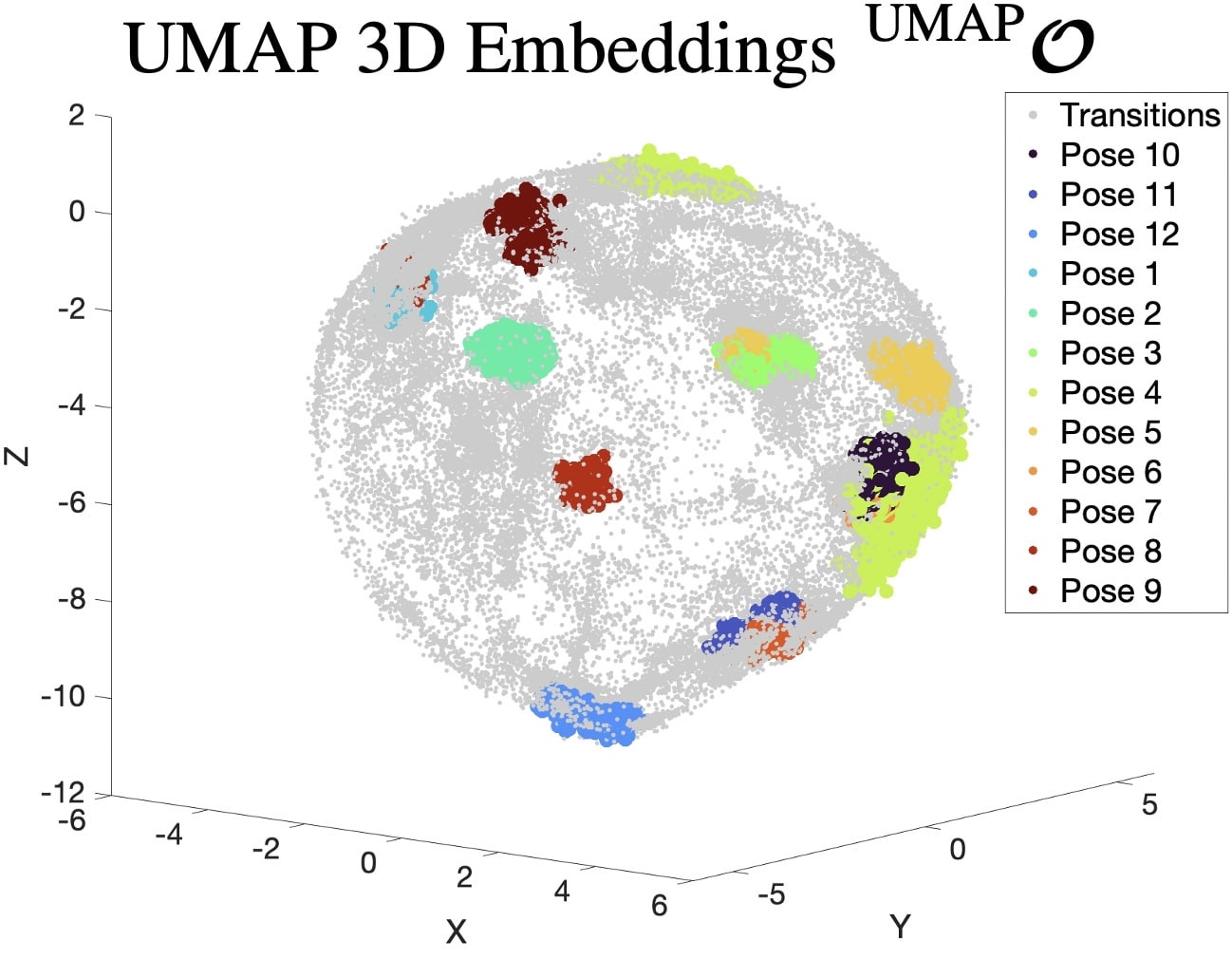}
    	\caption[Optional Caption]{}
    	\label{fig:DR_whole_recording_UMAP}
    \end{subfigure}
    \begin{subfigure}[b]{0.49\textwidth}
        \centering
    	\includegraphics[width=\linewidth]{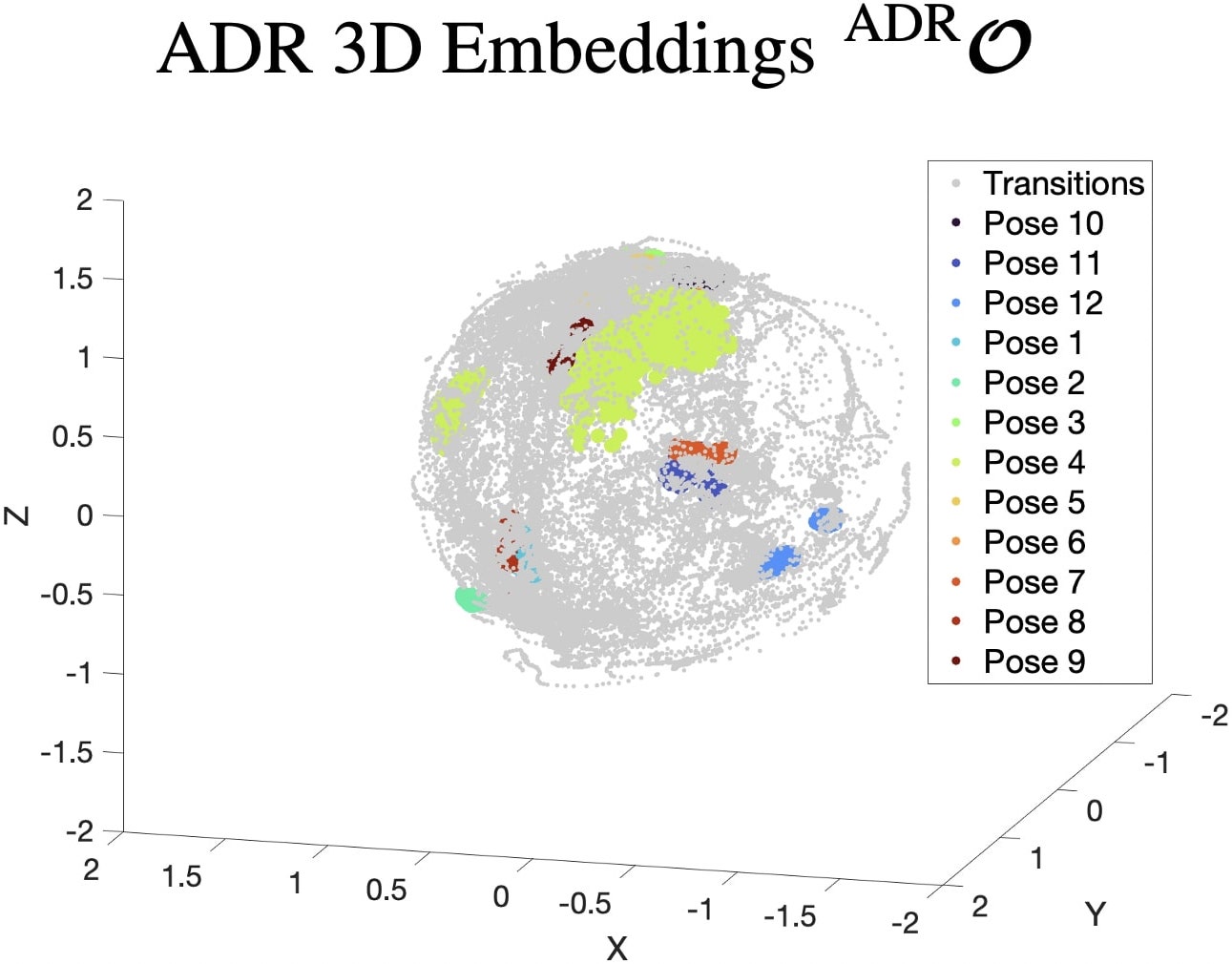}
    	\caption[Optional Caption]{}
    	\label{fig:DR_whole_recording_ADR}
    \end{subfigure}
    \par
    \begin{subfigure}[b]{1.0\textwidth}
        \centering
    	\includegraphics[width=\linewidth]{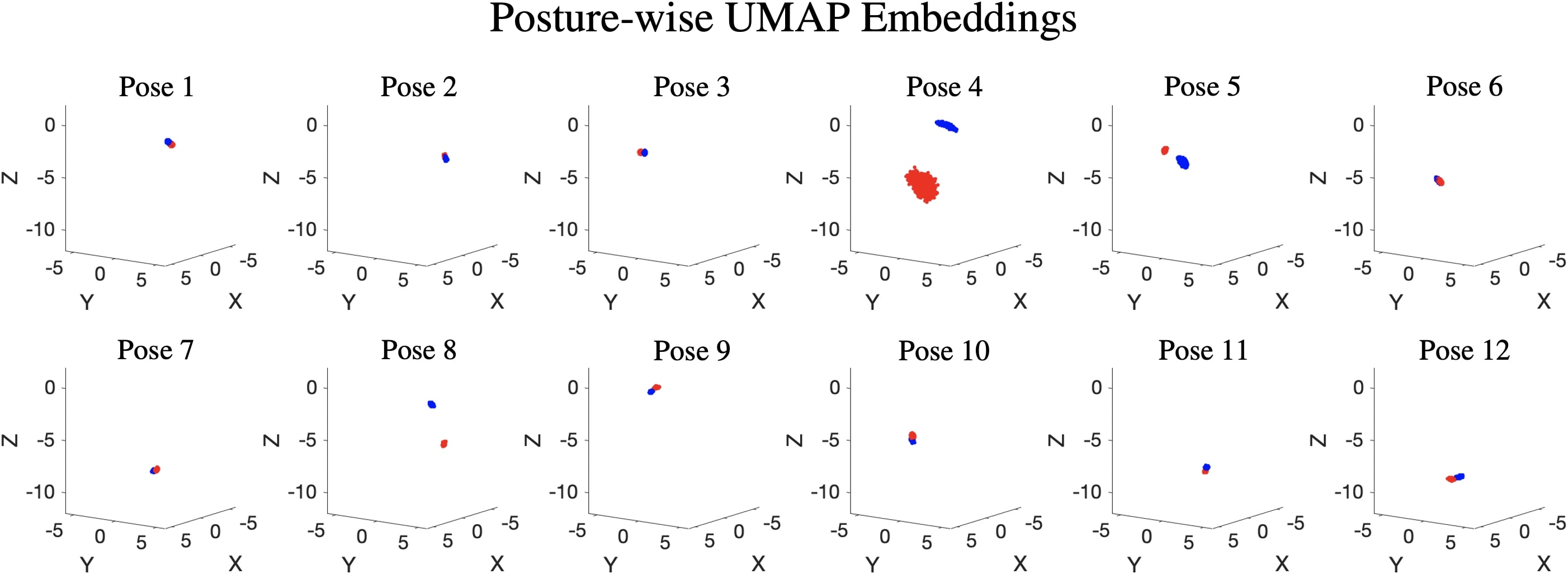}
    	\caption[Optional Caption]{}
    	\label{fig:UMAP_poses}
    \end{subfigure}
    \caption{Visualisation of a participant's wrist kinematics using UMAP and ADR techniques: (a) a 3D point cloud formed by all LW joint orientations observed in the datasets (postures and transitions are colour-coded), and (b) the blue- and red-coloured 3D UMAP embedding clusters produced from the dual replication of each sleep posture.}
    \label{fig:DR_LW_visualisations}
\end{figure}

{\color{black} A {\itshape computer graphics} (CG) pipeline (outlined in the Methods Section) was adopted to produce a synthetic dataset of nearly 50,000 axis-angle orientations, enabling UMAP to learn the 4D-to-3D mapping task without longitudinal collection of sensor data. In the 3D embedding space, UMAP represented the synthetic orientations dataset as a thick-crust, egg-shaped point cloud shown in \cref{fig:kids_framework}.}
Manual investigation showed that the latitudinal and longitudinal navigation of the 3D point cloud corresponded to different axes of rotation, whereas the radial distance from the cloud center was found proportional to the angle of rotation. Afterwards, the UMAP was presented with over 60 minutes of sensor-measured LW joint orientations from a random participant as depicted in \cref{fig:DR_whole_recording_UMAP}. From the figure it can be seen that the wrist orientations evidently occupy finite regions of the 3D embedding space while leaving some blank due to the anatomical joint constraints. Surprisingly, the twelve sleep postures are discriminable by the LW joint orientation alone (see \cref{fig:UMAP_poses}) except for minor overlaps which are typical.

The success of UMAP in the visualisation of joint kinematics does not necessarily lead to effective Bayesian inference (more on this will be discussed later). UMAP is intrinsically a stochastic method, meaning that different runs are not guaranteed to produce the same embedding space. This limitation can be partly addressed by saving the pre-trained UMAP model for later use. However, the unconstrained nature of the embedding space (non-origin centered, unevenly scaled and geometrically deformed point cloud) remain inevitable. Therefore, a second DR method was proposed to analytically approximate the nonlinear UMAP mapping function whilst giving full control over the embedding space. From here onwards, the latter method is regarded as {\itshape Analytical Dimension Reduction} (ADR).

Unlike the UMAP-based method, the proposed ADR does not require pre-training, and is designed to mathematically produce an origin-centered, thick-crust sphere with radius ranging from 1 to 2 in the Cartesian space. The latitudinal, longitudinal and radial displacements within the ADR embedding space have the same kinematic interpretations of that of the UMAP embedding space. For illustration purpose only, the synthetic dataset (designated for UMAP) was visualised using the ADR method as shown in \cref{fig:kids_framework} with no geometrical artefacts. \cref{fig:DR_whole_recording_ADR} shows the ADR-preprocessed LW joint orientations measured by the miniature wearable sensor from a random participant with similar observations in regards to anatomical constraints and discriminability of sleep postures.

\subsection*{(In)activity Detection and Segmentation}

The perception of physical (in)activity during sleep is probabilistically handled by Bayesian inference. The Bayesian inference operates on $\mathbfcal{O}$ after downsampling (decimation factor = $100$), indexed by a timestamp vector $\prescript{\Downarrow}{}{\bm t} = \prescript{\Downarrow}{}{t_1}, \prescript{\Downarrow}{}{t_2}, ..., \prescript{\Downarrow}{}{t_T}$. The KIDS framework capitalises on a Bayesian inference algorithm \cite{Adams2007} which evaluates weighted hypotheses on $\fancyR_k$ at each arbitrary time step, $\prescript{\Downarrow}{}{t_k}$ and sequentially estimates the posterior predictive parameters (e.g. mean and precision) of an input timeseries. The method presented in this paper reduces these weighted hypotheses to produce a single number which is the posterior mean estimate, $\hat{\fancyR}_k$, per time step $t_k$. A significant drop in this estimate implies that there is an increased probability of a short run length, i.e that the recent data likely belongs to a different distribution which, in turns, implies human subject activity. Further reset detection logic is applied to $\hat{\fancyR}_k$ to detect changepoint events and segment the periods of inactivity in the timeseries data. It was found that $\hat{\fancyR}_k$, though rarely, can exhibit a gradual reset with consecutive magnitude drops over two or three time steps; {\color{black} therefore, a postprocessing algorithm was applied to the run length estimate $(\hat{\fancyR}_k \rightarrow \hat{\fancyR}^p_k)$ to better detect reset events at both sudden and gradual falls. Complete details on (in)activity detection and segmentation are provided in the Methods Section.}

To realise the added value of each stage in the proposed methodology, four variations (two major and two minor) of the KIDS framework were evaluated. The two major variations come from the choice of the kinematic preprocessing method (either UMAP or ADR). As it will be shown later, the choice of the DR method does not only affect the topology of $\mathbfcal{O}$, but also substantially governs the level of information on $\mathbfcal{O}$ encoded in the initial prior presented to the Bayesian inference algorithm. The other two minor variations correspond to whether the postprocessing algorithm was used. Presented below are the detailed results for the best- and least-performing variations, respectively, of KIDS: (i) ADR with (w/) postprocessing and (ii) UMAP without (w/o) postprocessing. Nonetheless, the overall performance evaluation metrics are available for all four variations.

\subsubsection*{Analysis of ADR-based Bayesian Inference with Postprocessing (Best-performing Algorithm)}

\begin{figure}[!t]
    \centering
\begin{tikzpicture}
\node[above right] (PriorNonInfo) at (0,0) {\includegraphics[height=0.92\textheight]{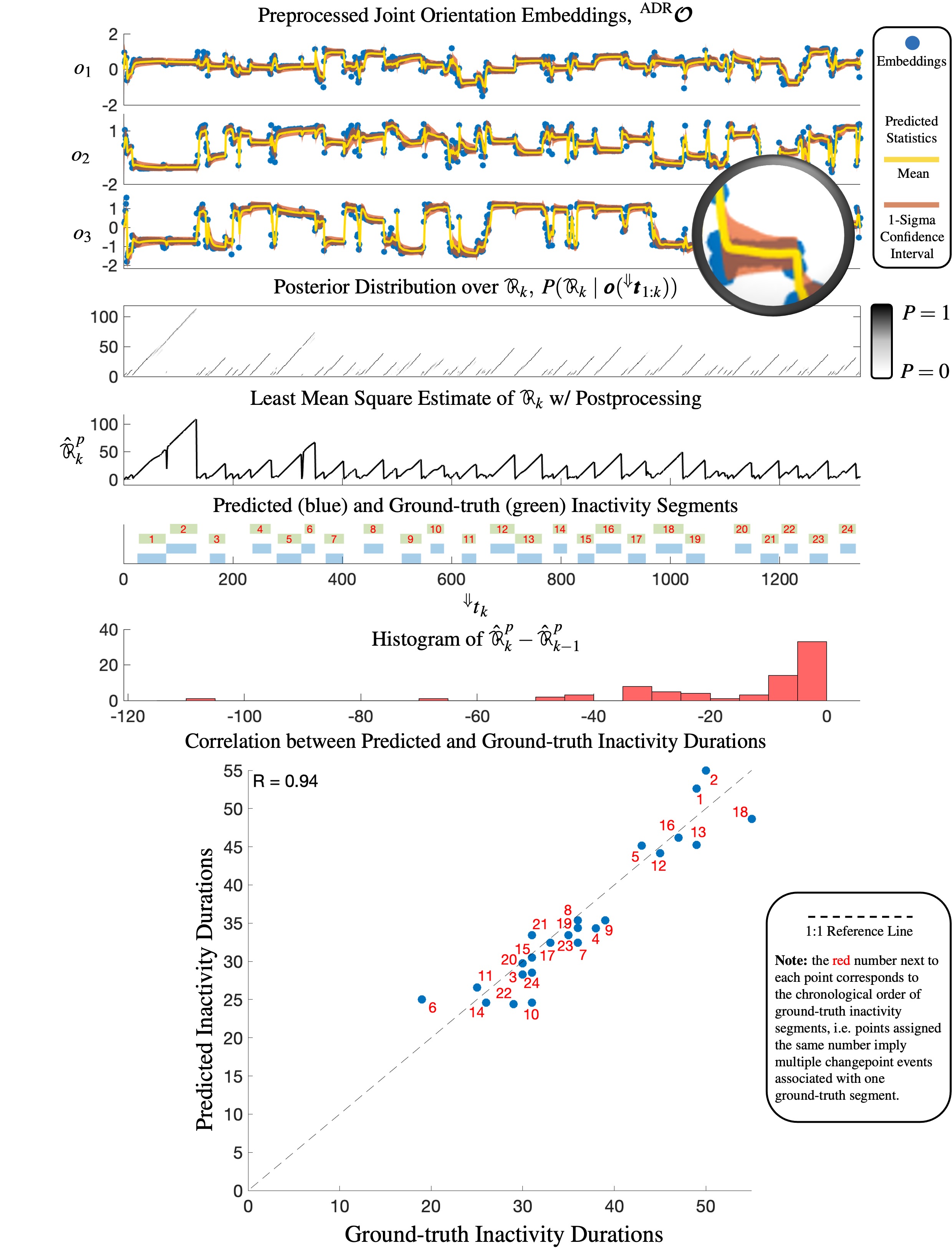}};


\draw[ultra thick,rounded corners,black] (-0.05\linewidth,0.80\textheight) rectangle (-0.05\linewidth+0.75cm,0.80\textheight+0.75cm) node[pos=.5,align=center] {\hypertarget{fig:ADR_w_PP_Results_a}{\large a}};

\draw[ultra thick,rounded corners,black] (-0.05\linewidth,0.655\textheight) rectangle (-0.05\linewidth+0.75cm,0.655\textheight+0.75cm) node[pos=.5,align=center] {\hypertarget{fig:ADR_w_PP_Results_b}{\large b}};

\draw[ultra thick,rounded corners,black] (-0.05\linewidth,0.58\textheight) rectangle (-0.05\linewidth+0.75cm,0.58\textheight+0.75cm) node[pos=.5,align=center] {\hypertarget{fig:ADR_w_PP_Results_c}{\large c}};

\draw[ultra thick,rounded corners,black] (-0.05\linewidth,0.505\textheight) rectangle (-0.05\linewidth+0.75cm,0.505\textheight+0.75cm) node[pos=.5,align=center] {\hypertarget{fig:ADR_w_PP_Results_d}{\large d}};

\draw[ultra thick,rounded corners,black] (-0.05\linewidth,0.42\textheight) rectangle (-0.05\linewidth+0.75cm,0.42\textheight+0.75cm) node[pos=.5,align=center] {\hypertarget{fig:ADR_w_PP_Results_e}{\large e}};

\draw[ultra thick,rounded corners,black] (-0.05\linewidth,0.19\textheight) rectangle (-0.05\linewidth+0.75cm,0.19\textheight+0.75cm) node[pos=.5,align=center] {\hypertarget{fig:ADR_w_PP_Results_f}{\large f}};

\end{tikzpicture}
\caption{Results from a participant dataset using ADR w/ postprocessing of $\hat{\protect\fancyR}_k$.
}
\label{fig:ADR_w_PP_Results}
\end{figure}
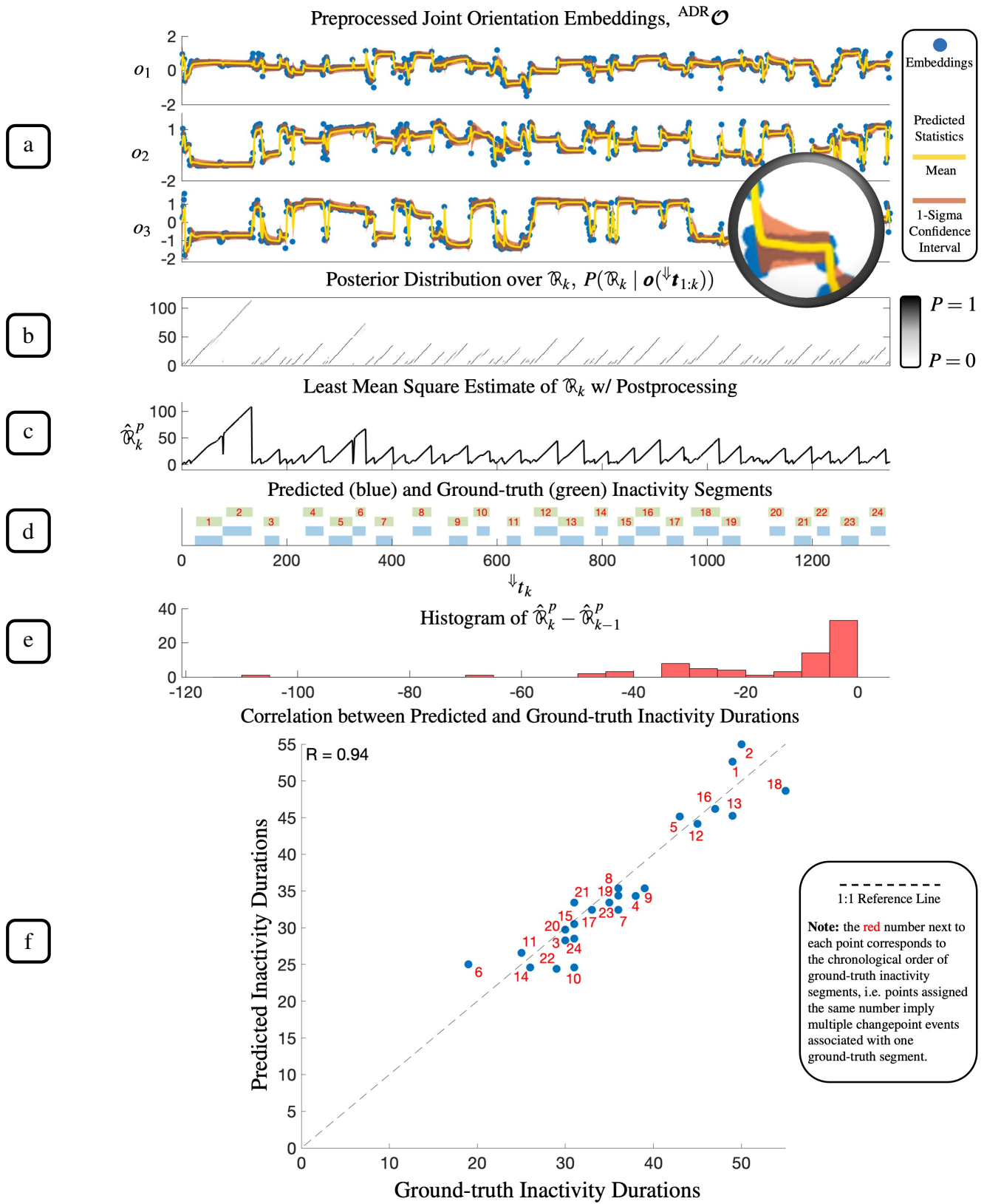

As discussed earlier, ADR produces a constrained 3D embedding space. Such controlled space advantageously facilitates crafting of the so-called ``{\itshape informative prior}'' (see \cref{fig:kids_framework}) to be presented to the Bayesian inference. The prior is typically encoded in the form of a parameterised probability distribution, describing the state of knowledge on $\prescript{\text{ADR}}{}{\mathbfcal{O}}$ before evidence measurements are obtained. {\color{black} More details are provided in the Methods section on the encoding of the informative prior belief.}

A random participant dataset was selected to {\color{black} comment on the (in)activity detection and segmentation results of KIDS shown in \cref{fig:ADR_w_PP_Results}.}
For this dataset, KIDS achieves a 100\% $F_1$-score in the changepoint detection task and a satisfactory correlation coefficient $(R=0.94)$ between predicted and ground-truth (GT) inactivity durations.
{\color{black} \cref{fig:ADR_w_PP_Results}\hyperlink{fig:ADR_w_PP_Results_a}{a} shows the 3D ${\bm o}$ embeddings along with the predicted current segment mean and standard deviation (by-products of Bayesian inference).}
The estimated statistics appear to effectively model the underlying (hidden) data sampling process unique to each segment of (in)activity. Upon the onset of each inactive segment, the 1-Sigma confidence interval (brown strip) gradually converges to the true underlying spread of the timeseries segment as shown in the exploded view given in \cref{fig:ADR_w_PP_Results}\hyperlink{fig:ADR_w_PP_Results_a}{a}.
The effectiveness of the segment-aware statistical modelling owes to the capability of the Bayesian inference to correctly assign probabilities (weights) to different hypotheses on $\fancyR_k$ {\color{black} as shown in the grayscale matrix representation of $P(\fancyR_k\ |\ {\bm o}(\prescript{\Downarrow}{}{t_{1:k}}))$ in \cref{fig:ADR_w_PP_Results}\hyperlink{fig:ADR_w_PP_Results_b}{b}.}


{\color{black} The KIDS framework employs multi-stage processing of $P(\fancyR_k\ |\ {\bm o}(\prescript{\Downarrow}{}{t_{1:k}}))$ to accomplish (in)activity detection and segmentation.}
First, a {\itshape Least Mean Square} (LMS) Bayesian estimator is used to elect one mean estimate, $\hat{\fancyR}_k$ from {\color{black} each column of} $P(\fancyR_k\ |\ {\bm o}(\prescript{\Downarrow}{}{t_{1:k}}))$.
{\color{black} Second, the point estimates of the run length are postprocessed, producing $\hat{\fancyR}^p_k$ shown in \cref{fig:ADR_w_PP_Results}\hyperlink{fig:ADR_w_PP_Results_c}{c}.}
{\color{black} Third, a reset detection logic (further explained in the Methods Section) then operates on $\log_{10}\hat{\fancyR}^p_k$ to detect the temporal locations of the changepoint events (end of inactivity). Fourth, combining the temporal locations of changepoint events with the point estimates of run length $\hat{\fancyR}^p_k$, the postural inactivity was segmented as shown in \cref{fig:ADR_w_PP_Results}\hyperlink{fig:ADR_w_PP_Results_d}{d}. A lower limit of 20 samples was imposed on the duration of segmented inactivity to eliminate incidental resets (false positives), which typicaly occur as the sleeper takes few transitions before settling on a new posture. The value of the lower limit was informed by the {\itshape multimodal histogram} (see \cref{fig:ADR_w_PP_Results}\hyperlink{fig:ADR_w_PP_Results_e}{e}) of magnitude drops in $\hat{\fancyR}^p_k$ at the temporal locations of changepoint events, where the rightmost peak corresponds to the aforementioned incidental resets.}
In rare cases where ${\bm o}$ embeddings are highly similar across two consecutive sleep postures (e.g. $1^\text{st}$ and $2^\text{nd}$ inactive segments in \cref{fig:ADR_w_PP_Results}\hyperlink{fig:ADR_w_PP_Results_d}{d}), it was found that $\hat{\fancyR}^p_k$ may exhibit a transient drop (partial reset) during the {\color{black} posture} transition.
{\color{black} If this occurs, the duration of the second inactive segment would be slightly overestimated. Therefore, an upper limit (= elapsed time in samples since the previous changepoint event) is imposed on the estimated inactivity duration.}
Portrayed in \cref{fig:ADR_w_PP_Results}\hyperlink{fig:ADR_w_PP_Results_f}{f} is the correlation plot between predicted and GT inactivity segment durations, indicating a satisfactory performance despite systematic errors due to signal discretisation and downsampling.



\subsubsection*{Analysis of UMAP-based Bayesian Inference without Postprocessing (Least-performing Algorithm)}

\begin{figure}[!t]
    \centering
\begin{tikzpicture}
\node[above right] (PriorNonInfo) at (0,0) {\includegraphics[height=0.92\textheight]{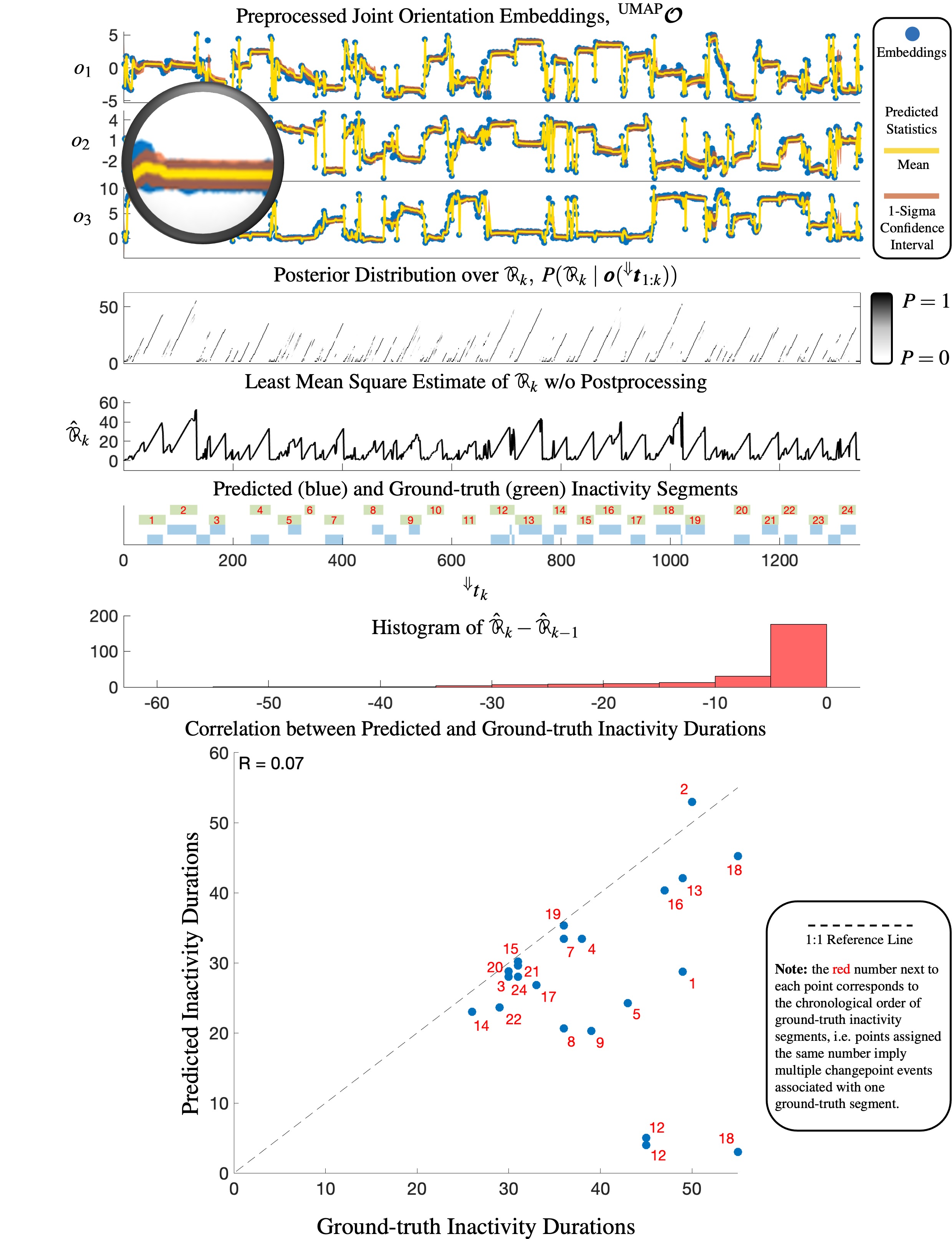}};


\draw[ultra thick,rounded corners,black] (-0.05\linewidth,0.80\textheight) rectangle (-0.05\linewidth+0.75cm,0.80\textheight+0.75cm) node[pos=.5,align=center] {\hypertarget{fig:UMAP_wo_PP_Results_a}{\large a}};

\draw[ultra thick,rounded corners,black] (-0.05\linewidth,0.655\textheight) rectangle (-0.05\linewidth+0.75cm,0.655\textheight+0.75cm) node[pos=.5,align=center] {\hypertarget{fig:UMAP_wo_PP_Results_b}{\large b}};

\draw[ultra thick,rounded corners,black] (-0.05\linewidth,0.59\textheight) rectangle (-0.05\linewidth+0.75cm,0.59\textheight+0.75cm) node[pos=.5,align=center] {\hypertarget{fig:UMAP_wo_PP_Results_c}{\large c}};

\draw[ultra thick,rounded corners,black] (-0.05\linewidth,0.515\textheight) rectangle (-0.05\linewidth+0.75cm,0.515\textheight+0.75cm) node[pos=.5,align=center] {\hypertarget{fig:UMAP_wo_PP_Results_d}{\large d}};

\draw[ultra thick,rounded corners,black] (-0.05\linewidth,0.42\textheight) rectangle (-0.05\linewidth+0.75cm,0.42\textheight+0.75cm) node[pos=.5,align=center] {\hypertarget{fig:UMAP_wo_PP_Results_e}{\large e}};

\draw[ultra thick,rounded corners,black] (-0.05\linewidth,0.20\textheight) rectangle (-0.05\linewidth+0.75cm,0.20\textheight+0.75cm) node[pos=.5,align=center] {\hypertarget{fig:UMAP_wo_PP_Results_f}{\large f}};

\end{tikzpicture}
\caption{Results from a participant dataset using UMAP w/o postprocessing of $\hat{\protect\fancyR}_k$.
}
\label{fig:UMAP_wo_PP_Results}
\end{figure}


Unlike ADR, UMAP produces an unconstrained 3D embedding space. Consequently, a ``{\itshape non-informative prior}'' is presented to the Bayesian inference (see \cref{fig:kids_framework}) to expand its ability of modelling $\prescript{\text{UMAP}}{}{\mathbfcal{O}}$ for which the data topology is ambiguous. As it will be further elaborated in the Methods section, the multivariate non-informative prior spreads widely (almost flat) over the {\itshape mean-covariance space} such that no particular combination is favoured in any way. This allows the Bayesian inference to objectively construct the posterior distribution prominently based on observed $\prescript{\text{UMAP}}{}{\mathbfcal{O}}$.

The same participant dataset was again used to comment on the {\color{black} (in)activity detection and segmentation} results of this variant of the KIDS framework{\color{black} , depicted in \cref{fig:UMAP_wo_PP_Results}}.
{\color{black} Overall, this variant showed poor performance. The changepoint detection $F_1$-score dropped to 76.9\% and the correlation coefficient (R) between predicted and GT inactivity durations dipped drastically to $7\mathrm{e}{-2}$.}

Due to the unconstrained pointcloud topology of UMAP, \cref{fig:UMAP_wo_PP_Results}\hyperlink{fig:UMAP_wo_PP_Results_a}{a} demonstrates that $o_1$, $o_2$ and $o_3$ had varying scales and offsets from zero. Despite it being the same participant dataset, $\prescript{\text{UMAP}}{}{\mathbfcal{O}}$ nonetheless exhibits more aggressive signal dynamics (large spikes) than $\prescript{\text{ADR}}{}{\mathbfcal{O}}$, specifically during posture transition times. This behaviour could simply be due to the stretched topology of the UMAP pointcloud (see \cref{fig:DR_whole_recording_UMAP}), or alternatively, it may suggest non-linearities (or discontinuities) in the UMAP mapping function though this would require further investigation to confirm. In regard to the current segment statistics, the Bayesian predicted mean and standard deviation demonstrate an underdamped dynamic response (see \cref{fig:UMAP_wo_PP_Results}\hyperlink{fig:UMAP_wo_PP_Results_a}{a}) which contrasts with the damped statistical predictions from the earlier KIDS variant in \cref{fig:ADR_w_PP_Results}\hyperlink{fig:ADR_w_PP_Results_a}{a}. This is due to the absence of an informative prior, forcing the Bayesian inference to largely rely on observed ${\bm o}$ to estimate the timeseries statistics. This observation is supported by the 1-Sigma confidence interval that fits tightly onto ${\bm o}$ embeddings (refer to the exploded view of \cref{fig:UMAP_wo_PP_Results}\hyperlink{fig:UMAP_wo_PP_Results_a}{a}), unlike the posterior statistics of the previous KIDS variant which takes only a few time steps to gradually converge from the prior covariance to the true underlying covariance of the timeseries.

The heavy dependency of Bayesian inference on ${\bm o}$ embeddings led to a higher uncertainty in the Bayesian multi-hypothesis evaluation of $\fancyR_k$. The uncertainty can be visually observed in \cref{fig:UMAP_wo_PP_Results}\hyperlink{fig:UMAP_wo_PP_Results_b}{b} in the form of ``{\itshape salt and pepper noise}'' contaminating the grayscale matrix representation of $P(\fancyR_k\ |\ {\bm o}(\prescript{\Downarrow}{}{t_{1:k}}))$. As a result, the LMS Bayesian estimator outputted a less reliable point estimate $\hat{\fancyR}_k$ (refer to \cref{fig:UMAP_wo_PP_Results}\hyperlink{fig:UMAP_wo_PP_Results_c}{c}) that was too sensitive to minor variations in $\prescript{\text{UMAP}}{}{\mathbfcal{O}}$. These minor variations are typically due to involuntary micro-body movements, such as hand twitches and breathing-related perturbations to the wrist pose.

{\color{black} A multi-stage processing of $P(\fancyR_k\ |\ {\bm o}(\prescript{\Downarrow}{}{t_{1:k}}))$, similar to that of the previous variant of KIDS, was performed except that no postprocessing was applied to the LMS point estimate $\hat{\fancyR}_k$. Three main observations can be concluded from the (in)activity detection and segmentation results shown in \cref{fig:UMAP_wo_PP_Results}\hyperlink{fig:UMAP_wo_PP_Results_d}{d}. First, numerous repetitive resets were detected (see \cref{fig:UMAP_wo_PP_Results}\hyperlink{fig:UMAP_wo_PP_Results_e}{e}), some of which were unfiltered incidental resets, explaining the presence of short-lasting inactive segments (see the {\color{red}12}\textsuperscript{th} and {\color{red}18}\textsuperscript{th} GT inactive segments). Second, few GT inactive segments ({\color{red}6}\textsuperscript{th}, {\color{red}10}\textsuperscript{th} and {\color{red}11}\textsuperscript{th}) were neither detected nor segmented. Third, the correlation plot in \cref{fig:UMAP_wo_PP_Results}\hyperlink{fig:UMAP_wo_PP_Results_f}{f} shows strong evidence of underestimation of inactivity durations, i.e. most segmentations are located below the 1:1 reference line. All three observations were primarily caused by the poor reliability of the run length estimation performed by this KIDS variant.}

\subsubsection*{Performance Metrics: A Comparative Analysis on Variants of the KIDS Framework}

The performance evaluation metrics are essential not only to make a fair comparison between the four variants of the proposed KIDS framework, but also to recognise the added value of each of the adopted methods, such as UMAP, ADR and postprocessing of $\hat{\fancyR}_k$. In this work, the {\itshape $F_1$-score}, {\itshape Sensitivity} (Se) and {\itshape Positive Predictive Value} (PPV) are the metrics used to evaluate the changepoint detection performance, while the {\itshape Pearson's correlation coefficient} (R) is reserved for assessing the quality of inactivity segmentation. These are standard metrics commonly reported in relevant works \cite{Palotti2019} and would establish a good ground for benchmarking.

\cref{eqn:eq1}, where TP, FP and FN denote true positives, false positives and false negatives respectively, shows how each of the changepoint detection metrics is determined. The predictive positive value (or precision) is the ratio of correct KIDS-derived changepoint events $(\prescript{\text{KIDS}}{}{n_c})$ to the total number of changepoint events $(\prescript{\text{KIDS}}{}{n})${\color{black}; a detected changepoint event is regarded as ``correct'' if it is no more than 3 samples away from the end time of its closest GT inactive segment.}
The sensitivity (or recall) is the ratio of $\prescript{\text{KIDS}}{}{n_c}$ to the total number of GT inactive segments $(\prescript{\text{GT}}{}{n})$. Lastly, the $F_1$-score is the harmonic mean of PPV and Se. The Pearson's R coefficient is a test statistic that measures the statistical relationship between two continuous variables. According to \cref{eqn:eq2}, R is mathematically defined as the ratio of the covariance for two arbitrary random variates, X and Y, to the product of their standard deviations, $\sigma_X$ and $\sigma_Y$. This ratio ranges from +1 (complete positive correlation) to -1 (complete negative correlation), with 0 indicating no correlation. In this work, the sample populations of X and Y are the predicted and GT inactive segment durations respectively.

\begin{equation}
\label{eqn:eq1}
\begin{split}
\text{PPV} &= \frac{\prescript{\text{KIDS}}{}{n_c}}{\prescript{\text{KIDS}}{}{n}} \left(=\frac{\text{TP}}{\text{TP + FP}}\right)\\
\text{Se} &= \frac{\prescript{\text{KIDS}}{}{n_c}}{\prescript{\text{GT}}{}{n}} \left(=\frac{\text{TP}}{\text{TP + FN}}\right)\\
F_1\text{-score} &= \frac{2 \times \text{PPV} \times \text{Se}}{\text{PPV + Se}}
\end{split}
\end{equation}

\begin{equation}
\label{eqn:eq2}
R = \frac{Cov(X,Y)}{\sigma_X \cdot \sigma_Y}
\end{equation}


\begin{figure}[!t]
    \centering
    
    \begin{subfigure}[b]{0.49\linewidth}
        \centering
        \includegraphics[width=\textwidth]{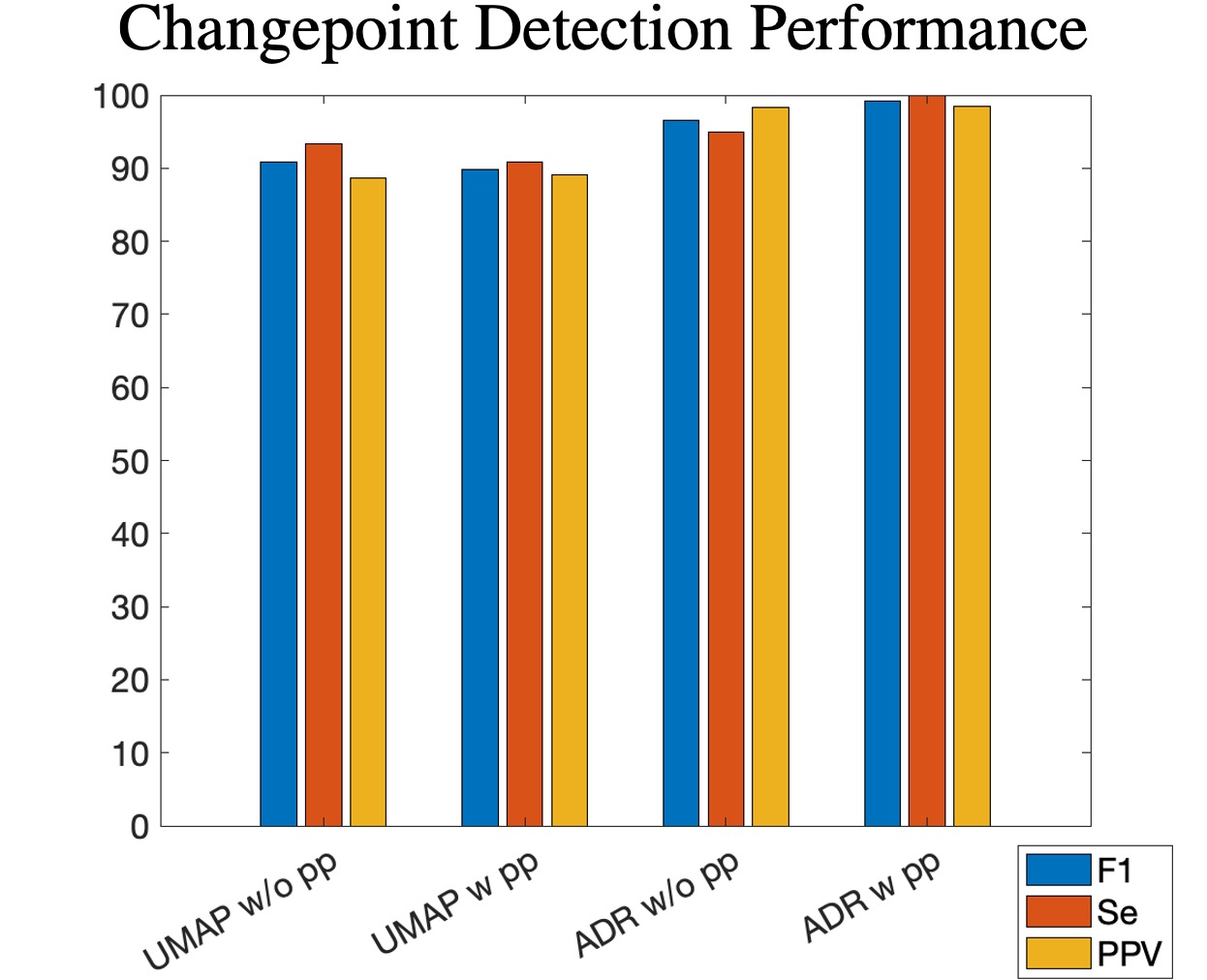}
        \caption{}
        \label{fig:ChangepointDetectionMetrics}
    \end{subfigure}
    \hfill
    \begin{subfigure}[b]{0.49\linewidth}
        \centering
        \includegraphics[width=\textwidth]{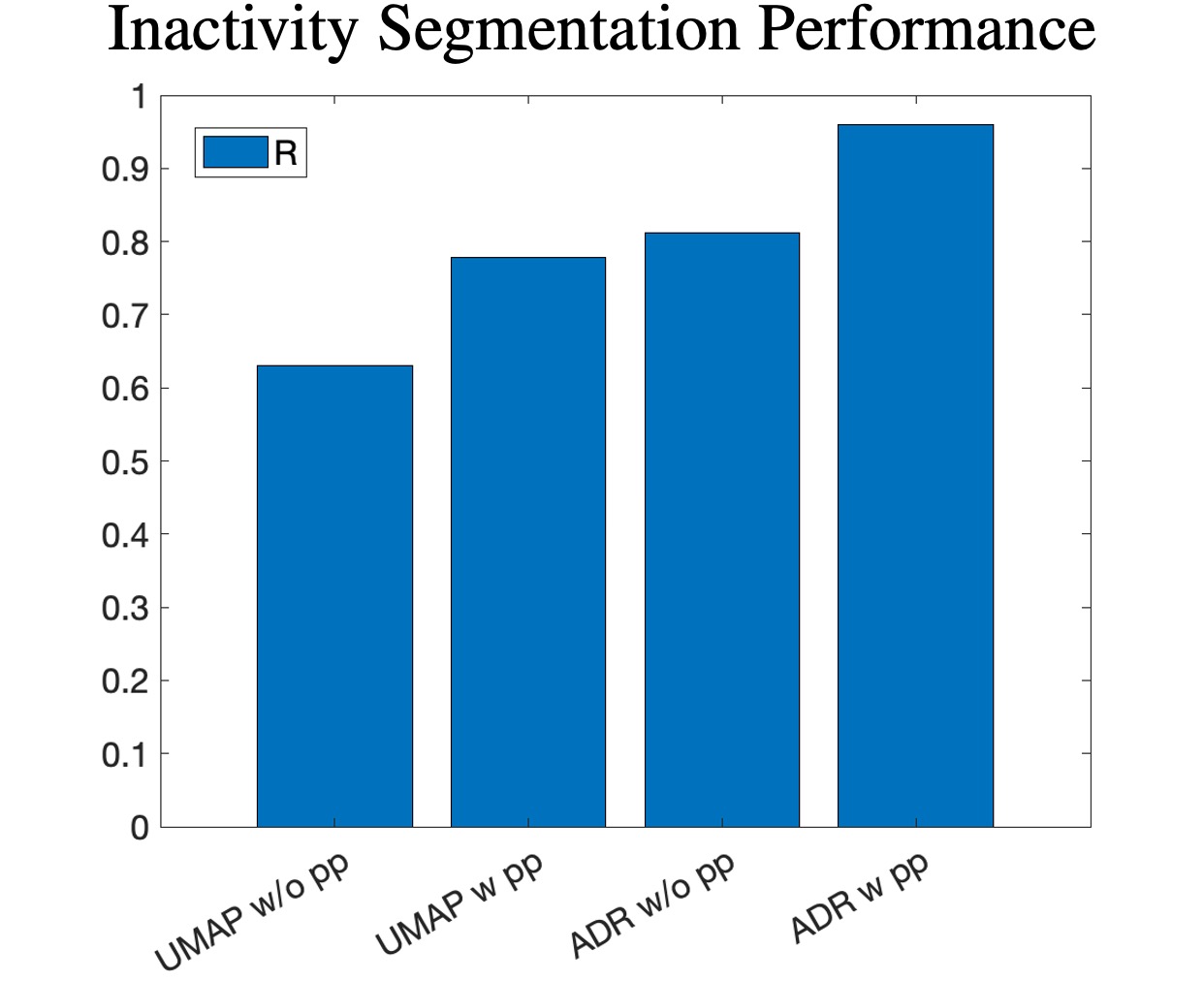}
        \caption{}
        \label{fig:InactivitySegmentationMetrics}
    \end{subfigure}
    
    \caption{Performance evaluation metrics of all four variants of the proposed KIDS framework.}
    \label{fig:PerformanceMetrics}
\end{figure}

For the changepoint detection performance (see \cref{fig:ChangepointDetectionMetrics}), ADR-based KIDS variants attained +96\% $F_1$ scores while UMAP-based variants scored below 91\% for the same metric.
{\color{black} The >5\% gap in performance underscores the positive impact brought by both the ADR 3D embedding space and the informative Bayesian prior to the detection of changepoint events.}
The postprocessing of $\hat{\fancyR}_k$ evidently improved the changepoint detection performance in the ADR case by up to 6.7\%, 12.5\% and 4.3\% gains in the $F_1$-score, Se and PPV metrics respectively. Importantly, the employed postprocessing algorithm stands out from the conventional re-scoring rules in the literature which are often criticised for compromising the $F_1$-score and Se in favour of the overall accuracy \cite{Palotti2019}. This is because existing re-scoring rules aim to improve the performance of an algorithm using assumptions on the sleeper, unlike the case in this work where the postprocessing algorithm was crafted based on the studied behaviour of the KIDS framework itself. Nevertheless, in the case of UMAP-based KIDS, postprocessing was found to have no to minor adverse effect on their changepoint detection performance, with an average $1\%$ drop in the $F_1$-score after the incorporation of the postprocessing algorithm. The unviability of $\hat{\fancyR}_k$ postprocessing for UMAP-based KIDS comes as no surprise, since the main problem was the intrinsic Bayesian uncertainty (due to missing an informative prior).

In regard to the inactivity segmentation performance (refer to \cref{fig:InactivitySegmentationMetrics}), the average difference in the Pearson's R coefficient between respective ADR- and UMAP-based KIDS variants is 0.18. The wide gap again signifies the importance of the ADR 3D embedding space and the informative prior in Bayesian inference. Besides, the postprocessing algorithm brought a surge in segmentation performance, with an average 23.5\% and 18.2\% enhancement to the Pearson's R coefficient of UMAP- and ADR-based KIDS respectively. Notably, the best-performing KIDS variant (ADR w/ postprocessing) accomplished an average R of 0.96, which is far more superior than the least-performing variant (UMAP w/o postprocessing) at around 0.63 only.

The key takeaways from the comparative analysis given in \cref{fig:PerformanceMetrics} are as follows. First, ADR-based KIDS variants generally perform better than UMAP-based variants at both changepoint detection and inactivity segmentation tasks, regardless of postprocessing $\hat{\fancyR}_k$. Second, when the postprocessing algorithm was factored into the performance evaluation, its effect was different for ADR- and UMAP-based KIDS. In case of ADR, the postprocessing algorithm was found to always enhance the framework performance in both tasks of changepoint detection and inactivity segmentation. However, for UMAP-based variants, postprocessing was only viable at the inactivity segmentation task for reasons covered earlier.

\section*{Discussion}

This paper has presented KIDS, which according to the authors' best knowledge, is the first in-bed movement analysis framework to fulfil two intertwined clinical needs: (in)activity detection and segmentation. Unlike the previous work reported in the literature that typically addresses either one of the two needs, the KIDS framework reformulates the problem of in-bed movement analysis by providing a joint answer to the two interconnected research questions. Moreover, while the prevailing literature used hard-coded thresholding of raw/processed sensor data, data-hungry classifiers or a mixture of the two approaches; KIDS leverages Bayesian probability at its core to provide an objective assessment of body (in)activity. The input to KIDS is wearable sensor measurements of the underlying kinematic profile of the left wrist joint. Such information is clinically meaningful and comprehensible to medical staff, and its use cases can be expanded beyond (in)activity analysis.

The KIDS framework complements a previous study \cite{Elnaggar2022a} on sleep posture recognition where four wearable sensor modules mounted on the wrists and ankles were used to classify twelve postures based on quadruple joint kinematic cues. For that purpose, four joints were monitored to minimise the likelihood of overlap between sleep postures sharing similar extremity limb positions. However, for the objective of (in)activity detection and segmentation during sleep, monitoring quadruple joints is not well justified. The hand is probably one of the most moved parts of the human body, and being lightweight, it potentially carries much of the information on body mobility during sleep. Therefore, as a starting point, this paper tested the hypothesis that monitoring the left wrist alone would be sufficient to obtain reliable performance.

Low power consumption and real-time performance are desired criteria for portable sleep monitoring devices.
The choice of the sensor data rate directly determines the update rate of the in-device algorithm, and power consumption consequently varies depending on the computational cost of the algorithm. As covered in the Introduction, it is a predominant practice in the literature to make use of actigraphy signals for wearable-based sleep analysis. While actigraphy captures higher-order kinematics of the human body, it typically necessitates a relatively fast-updating movement analysis algorithm in order not to miss posture transitions. Alternatively, the KIDS framework employs a computationally efficient inertial sensor fusion algorithm to fuse high-frequency inertial signals and produce a filtered estimate of the LW joint orientation that is robust to sensor artefacts and environmental disturbances. KIDS lays down the assumption that upon every posture transition, the LW joint orientation is permanently changed, meaning that the requirement of a fast-updating algorithm can be dropped. Therefore, downsampling of the joint orientation profile was safely incorporated into KIDS without concerns over the (in)activity detection and segmentation performance. As a result, the tight time constraints on the computation cycle of the in-device algorithm were substantially relaxed, allowing for the adoption of more advanced body movement analysis approaches such as the Bayesian probabilistic framework adopted in KIDS without compromising real-time performance.

For further algorithm explainability and more optimal real-time performance, KIDS incorporates a novel joint kinematics preprocessing stage based on dimension reduction to learn a low-dimensional representation of joint orientations. The LW joint kinematics recordings obtained from in-vivo experiments were embedded into intuitive 3D visualisations showing a decent intra-posture clustering and inter-posture separability. Visualisations produced by two DR methods (UMAP and ADR) were compared, and fully constrained 3D embedding spaces were only guaranteed by ADR. A side benefit of the constrained ADR 3D embedding space was that it facilitated the design of informative priors useful for Bayesian inference.

The general working principle of KIDS relies on statistical modelling of the 3D preprocessed joint kinematic timeseries to evaluate multiple hypotheses on the current segment run length. The output weighted hypotheses are then fused and processed by a reset detection logic to perform (in)activity detection and segmentation. The performance of four variants of the KIDS framework was quantitatively and qualitatively studied and comparisons were made based on standard metrics. The primary variants of KIDS come from the choice of the DR method used for preprocessing joint kinematics, while secondary variants correspond to whether the current segment run length estimate was refined by a postprocessing algorithm. It was found that ADR-based KIDS variants generally outperformed UMAP-based variants. Furthermore, even though the postprocessing algorithm enhanced the segmentation performance of UMAP-based KIDS, its added value was more evident in case of ADR-based KIDS improving both its changepoint detection and segmentation. Lastly, performance evaluation showed that the assumption of KIDS on permanent change in joint orientation upon each posture transition is valid except for only one case. For this case, the LW joint orientation of a participant barely changed after one posture transition; nonetheless, the timeseries modelling of KIDS was still able to partially reset $\hat{\fancyR}_k$ and the transition was successfully detected by the reset detection logic.

\section*{Methods}

{\color{black} This section presents the methodology and experimental protocol behind the KIDS framework. First, we discuss the participant study pipeline and the simulated sleep protocol devised for the validation of the framework. Second, an overview on the custom-made wearable sensors and the algorithm for wrist kinematics measurement is provided. Third, we outline the proposed DR-based preprocessing of measured joint kinematics using two methods; UMAP and Analytical dimension reduction. Lastly, the Bayesian probabilistic framework enabling the (in)activity detection and segmentation is outlined.}

\subsection*{Participant Study}

Five healthy adult participants  (age: 36 $\pm$ 15.8 years, height: 169 $\pm$ 11 cm, body weight: 72.8 $\pm$ 23.2 kg) took part in the study upon signing an informed consent. The methods were carried out in accordance with relevant guidelines and regulations and all experimental protocols were approved by The University of Liverpool Research Ethics Committee (review reference: 9850). A leaflet containing pictures of twelve sleep postures (refer to previous work \cite{Elnaggar2022a} for posture definitions) was handed to each participant to assist them in replicating the postures, with each sleep posture replicated twice (two trials). To ensure postural data resembles that of a realistic sleep scenario, a random pose shuffling technique was used to ensure statistical independence of samples across the dataset.
{\color{black} Posture and transition durations may vary as per the participant's comfort and need for guidance by on-site researchers.} For all datasets, a bespoke wearable sensor module was used to capture body segment and joint kinematics, and to transmit these data to a localhost server {\color{black}at a data rate of 30 Hz}.

\subsection*{Bespoke Wearable Sensor Module}

The participant study employed four wearable sensor modules to monitor the joint orientations of the two wrists and two ankles simultaneously. In a previous work \cite{Elnaggar2022a}, the multi-sensor data were suitable for a robust sleep posture classification. This work however exploits the LW joint kinematics alone, as a starting point, for body (in)activity analysis. The custom-made sensor module provides dual-segment orientation tracking across the LW joint, empowered by two embedded IMU sensors mounted on the forearm and hand. The IMU model is the BNO055 from Bosch Sensortec\copyright\ (Bosch Sensortec GmbH, Reutlingen, DE). Both IMU sensors are managed by a single ESP32-WROOM-32D microcontroller from Espressif Systems\copyright\ (Espressif Systems Shanghai Co Ltd, Shanghai, CN) featuring Bluetooth connectivity for wireless data transmission. At about 6 cubic centimeters in volume for each IMU case, the sensor module is sufficiently slim and small for wearability during sleep.

\subsection*{Inertial Sensor Fusion for Wrist Kinematics Measurement}

{\itshape Intra- and inter-sensor fusion} were employed in the work. An inertial sensor fusion algorithm was needed to estimate the attitude of each IMU sensor (intra-sensor fusion), that is, a function of the body segment it is mounted on. Given the two segment orientations derived from both IMU sensors, an inter-sensor fusion step is subsequently applied to determine the relative joint orientation. Prior to each in-vivo experiment, all IMU sensors were calibrated according to standard procedures \cite{Woodman2007,Kok2017} to estimate and reduce errors owing to constant bias, scale factors, cross-axis sensitivity and response nonlinearity.

For intra-sensor fusion, the Madgwick filter \cite{Madgwick2011} is employed to fuse the IMU geo-inertial measurements and provide a filtered estimate of the absolute segment quaternion with respect to the Earth reference frame. The filter has low computational cost and operates in the quaternion space, allowing for efficient and singularity-free IMU attitude estimation. For inter-sensor fusion, the relative segment-to-segment quaternion was found through kinematic transformation which maps the orientation of the child segment (hand) to that of the parent segment (forearm). For further details on the mathematical formulations surrounding sensor fusion, it is advised to refer to previous work \cite{Elnaggar2022a}. Lastly, the segment-to-segment orientation was then converted from the quaternion space to the axis-angle space, where a unique ${\bm x}(t_k) \in \mathbb{R}^4$ exists at each arbitrary time step $t_k$. The disentangled axis-angle representation of joint orientations is more intuitive compared to quaternions and allowed for the extraction of meaningful postural analytics as shown in previous work \cite{Elnaggar2022a}.

\subsection*{Joint Kinematics Preprocessing}

The preprocessing of joint kinematics was enabled by dimension reduction, allowing for intuitive 3D visualisations of sensor-measured joint orientations and computationally efficient Bayesian inference subsequently.
Such 3D visualisation of joint kinematic timeseries facilitated the design of the informative prior which was found to enhance the Bayesian inference, and could potentially be useful for medical screening and/or diagnosis.
Dimension reduction has been successfully applied to visualise data across different domains, from wearable sensing \cite{Elnaggar2022a} through speech processing \cite{Elnaggar2019b} to knowledge exchange \cite{Elnaggar2021}.
The following text elaborates on the two DR methods employed in this paper.

\subsubsection*{Uniform Manifold Approximation and Projection}

UMAP \cite{McInnes2018} is a nonlinear DR method that iteratively constructs a lower dimensional {\itshape force-directed graph} representative of some high-dimensional dataset.
It is capable of handling nonlinear data manifolds and is known for preserving both the local and global data structures in the low-dimensional embedding space. In principle, UMAP performs dimension reduction over two stages: (i) identifying nearest neighbours and constructing a neighbour graph, and (ii) learning a low-dimensional representation through iterative minimisation of a cost function. Further details on UMAP can be found in its original paper \cite{McInnes2018}.

The use of UMAP in this work is distinct from the common use of DR in human motion analysis works \cite{Airaksinen2020,Zebin2018,Makela2022,Hamad2019}. Herein UMAP was run twice; a first run to learn the mapping from 4D joint orientation data to a 3D manifold using synthetic data and a second run to embed sensor-measured joint orientations (unseen data) into the pretrained 3D embedding space.
The synthetic dataset was carefully designed to uniformly sample the axis-angle orientation space (given some resolution), hence enabling UMAP to construct a reliable neighbour graph.
Afterwards, the 4D sensor-measured joint orientations were passed to UMAP to be projected into the pretrained 3D embedding space accordingly. The 3D visualisations of synthetic and real joint orientations are presented in \cref{fig:kids_framework,fig:DR_whole_recording_UMAP} respectively.

\begin{figure}[!t]
    \centering
    \includegraphics[width=0.8\textwidth]{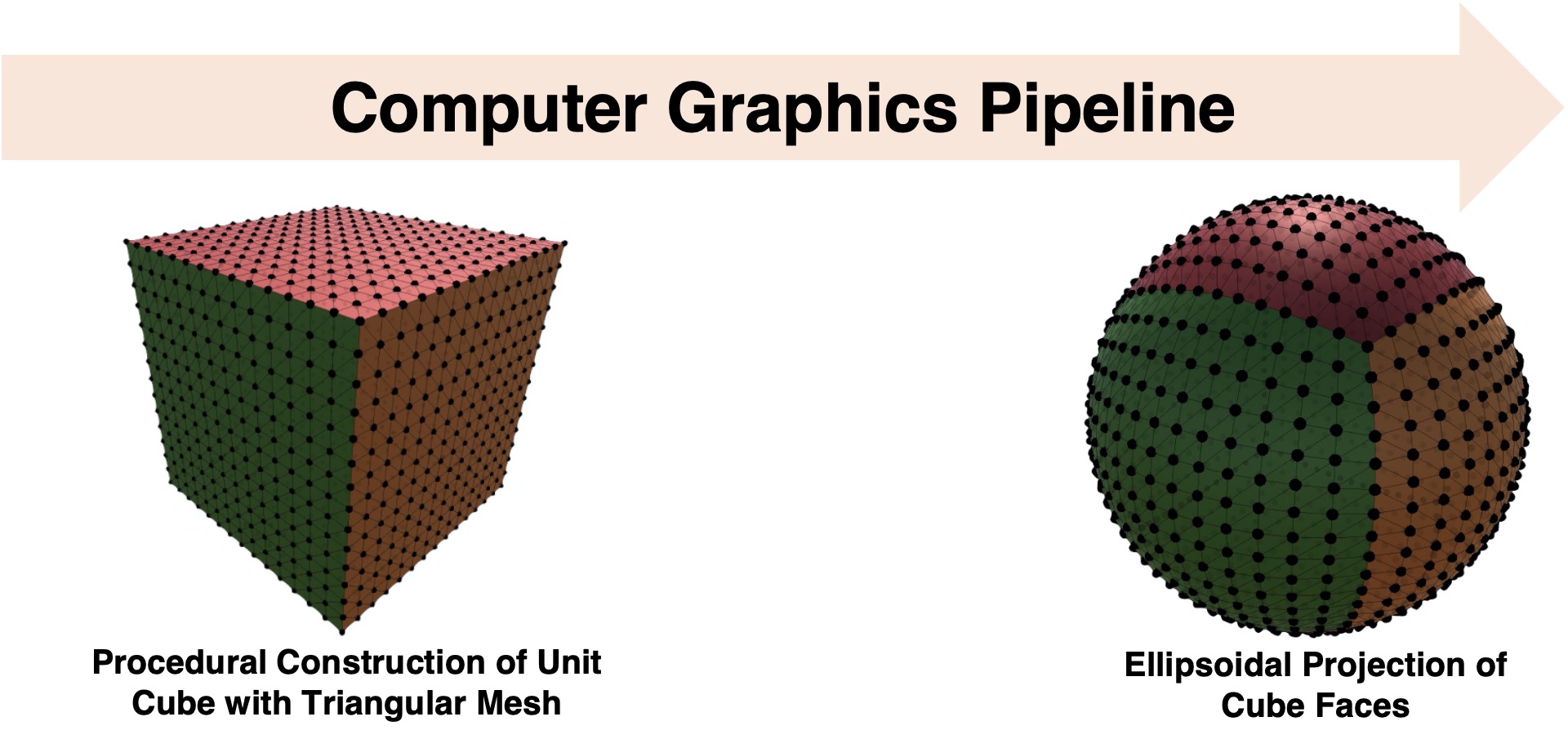}
    \caption{\color{black}The computer graphics pipeline used for the generation of synthetic axes of rotation.}
    \label{fig:CG_pipeline_ellipsoidal_projection}
\end{figure}

{\color{black}
A {\itshape computer graphics} (CG) pipeline (shown in \cref{fig:CG_pipeline_ellipsoidal_projection}) was used to generate the synthetic axes of rotation.
The pipeline first employed a {\itshape procedural mesh generation} technique to construct a unit cube with face vertices, ${\bm v}_f = \{x_f, y_f, z_f\}$. Then, the ellipsoidal projection
\begin{equation}
\label{eqn:eq10_manuscript}
{\bm v}_e
=
\begin{bmatrix}
x_e &
y_e &
z_e 
\end{bmatrix}
=
\begin{bmatrix}
\ x_f\sqrt{1 - \frac{1}{2}\ y_f^2 - \frac{1}{2}\ z_f^2 + \frac{1}{3}\ y_f^2z_f^2} &
y_f\sqrt{1 - \frac{1}{2}\ x_f^2 - \frac{1}{2}\ z_f^2 + \frac{1}{3}\ x_f^2z_f^2} &
z_f\sqrt{1 - \frac{1}{2}\ x_f^2 - \frac{1}{2}\ y_f^2 + \frac{1}{3}\ x_f^2y_f^2}\ 
\end{bmatrix}
\end{equation} was used to project all ${\bm v}_f$ onto the surface of a unit sphere, such that the projections are evenly distributed over the surface of the sphere. The projected vertices, ${\bm v}_e$, represent the synthetic axes of rotation, and were exported into a {\itshape comma-separated values} (CSV) file. All procedural 3D modelling was implemented in the C\# programming language and realised in Unity$^\copyright$ (Unity Technologies Inc., California, US).

The CSV file containing all ${\bm v}_e$ was subsequently imported into MATLAB$^\copyright$ (The MathWorks, Massachusetts, US). Therein, an orientation dataset generator script concatenated each synthetic axis of rotation with each angle of rotation from the closed set $\{\frac{36}{36}\pi,\frac{35}{36}\pi,\frac{34}{36}\pi,...,\frac{1}{36}\pi\}$. The output synthetic axis-angle dataset had a total of 48,600 orientations. Further details on the CG pipeline is available online in Supplementary Methods - CG Pipeline.
}

\subsubsection*{Analytical Dimension Reduction}

The 3D point cloud from UMAP resembled the shape a thick-crust egg shell. Through visual inspection, this point cloud was found not to be centred at the origin, and had geometrical artefacts such as uneven scaling. These are inevitable outcomes inherent to the unconstrained embedding space of UMAP. To gain full control over the dimension reduction as a process, {\itshape analytical dimension reduction} was proposed taking inspiration from the UMAP 3D visualisation.
In a nutshell, ADR aims to output a standardised point cloud taking the shape of a thick-crust sphere that is clear from any form of geometrical deformation (see \cref{fig:kids_framework}).

The ADR sphere have the following shape features. The sphere has an inner radius, $\prescript{i}{}{r_s} = 1$, and an outer radius, $\prescript{o}{}{r_s} = 2$. The radial displacement from $\prescript{i}{}{r_s}$ to $\prescript{o}{}{r_s}$ corresponds to a change in the angle of rotation from 0 to $\pi$ radians. Based on this definition, it is then possible to interpolate the radial displacement, $r_s$, given the sensor-measured angle of rotation $x_4$:
\begin{equation}
\label{eqn:eq11}
r_s = \frac{1}{\pi} x_4 + \prescript{i}{}{r_s}
\end{equation}

Additionally, the latitudinal and longitudinal displacements around the ADR sphere correspond to differently oriented axes of rotation defined by $x_1$, $x_2$ and $x_3$. Therefore, an arbitrary preprocessed orientation embedding, $\prescript{\text{ADR}}{}{\bm o}$, in the constrained embedding space of ADR can be formulated as
\begin{equation}
\label{eqn:eq12}
\prescript{\text{ADR}}{}{\bm o} =
\begin{bmatrix}
o_1 &
o_2 &
o_3 
\end{bmatrix}
= r_s
\begin{bmatrix}
x_1 &
x_2 &
x_3 
\end{bmatrix}
\end{equation}

\subsection*{Downsampling of Preprocessed Joint Kinematics}

Sleep time is dominated by long durations of inactivity. Running Bayesian inference at every time step is a waste of computational energy, and is not recommended for a wearable device that continuously streams sensor data at 30 Hz overnight. In such case, downsampling is generally a good solution to reduce the computational and data storage requirements. Therefore, the preprocessed joint kinematic timeseries was downsampled by a decimation factor of $100$ before it was presented to the Bayesian inference.

\subsection*{Bayesian Inference of the Current Segment Run Length}

\begin{figure}[!t]
    \centering
    \includegraphics[height=0.93\textheight]{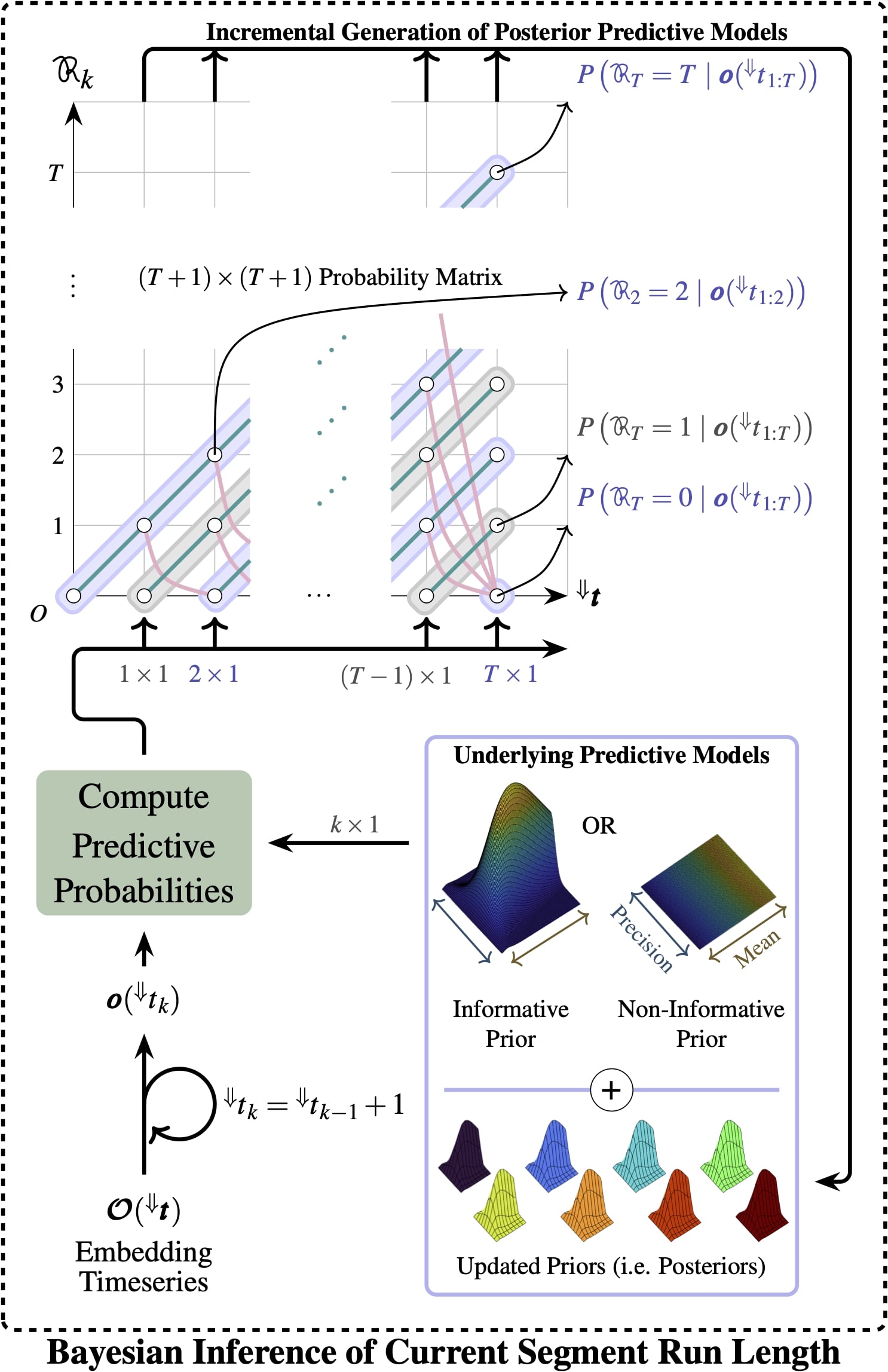}
    \caption{Graphical illustration of the Bayesian run length estimation algorithm.}
    \label{fig:bocpdDiagram}
\end{figure}

A prominent highlight of the KIDS framework is the use of Bayesian inference for (in)activity detection and segmentation. The Bayesian inference was used to estimate the current segment run length, $\fancyR_k$, at each time step $\prescript{\Downarrow}{}{t_k}$. By definition, $\fancyR_k$ is the length (in samples) of a data segment whose samples shares similar statistical characteristics. Essentially, $\fancyR_k$ is a key quantity which carries information on the length of inactivity and the temporal locations of changepoints.

The Bayesian inference in this paper is largely inspired by a previously reported Bayesian run length estimation algorithm \cite{Adams2007}, which evaluates multiple probability-weighted hypotheses on $\fancyR_k$ at each time step.
The original paper presented the general Bayesian framework for the estimation of $\fancyR_k$. Therefore, the following steps were incorporated to adapt the Bayesian framework to the case study presented in this paper.
First, the Bayesian inference problem was reformulated to simultaneously estimate both the mean and precision of the preprocessed joint kinematic timeseries. The estimation of the mean was essential since the preprocessed timeseries typical exhibit a mean shift upon a change in posture. The estimation of the precision was equally important since the fluctuations in the joint orientation slightly varies from one posture to another due to factors, such as snoring and breathing. Second, conjugate exponential distributions were adapted to the dual mean-precision estimation problem to maintain real-time inference. Third, a Bayesian estimator was utilised to fuse the weighted hypotheses on $\fancyR_k$ and output a single point estimate, $\hat{\fancyR}_k$ at each time step. Fourth, we complemented the original Bayesian framework with a {\itshape reset detection logic}, which detected the temporal locations of changepoints based on the $\hat{\fancyR}_k$ estimate.

The following section explains the general working principle behind the Bayesian inference framework.
{\color{black} Complete details on the underlying mathematics are available online in Supplementary Methods - Bayesian Estimation of Current Segment Run Length, including the construction of the prior probability distribution used for implicit encoding of information \cite{Murphy2007} to tune the Bayesian inference.}

\subsubsection*{General Overview of the Bayesian Run Length Estimation}

A graphical illustration of the Bayesian run length estimation algorithm is presented in \cref{fig:bocpdDiagram}. The algorithm requires two pieces of information as inputs: (i) either informative or non-informative prior on $\mathbfcal{O}(\prescript{\Downarrow}{}{\bm t})$, and (ii) the embedding timeseries $\mathbfcal{O}(\prescript{\Downarrow}{}{\bm t})$. Initially at $\prescript{\Downarrow}{}{t_0}$ and before observing $\mathbfcal{O}$, the algorithm assumes $P(\fancyR_0 = 0,\ {\bm o}(\prescript{\Downarrow}{}{t_0}) = \phi) = P(\fancyR_0 = 0) = 1$ which is the probabilistic notation for $\fancyR_0 = 0$. At $\prescript{\Downarrow}{}{t_1}$ and upon observing ${\bm o}(\prescript{\Downarrow}{}{t_1})$, the algorithm evaluates the {\itshape predictive probability}; the probability that ${\bm o}(\prescript{\Downarrow}{}{t_1})$ belong to the (non-)informative prior.
Afterwards, the algorithm evaluates all possible hypotheses on $\fancyR_1$. Logically, $\fancyR_k$ can  take one of two values:
\begin{equation}
\label{eqn:eq13}
\fancyR_k =
\begin{cases}
0, & \text{if changepoint occurs at $\prescript{\Downarrow}{}{t_k}$}.\\
\fancyR_{k-1} + 1, & \text{otherwise}.
\end{cases}
\end{equation}

Nonetheless, the Bayesian framework does not directly set $\fancyR_k$. Instead, it evaluates two types of transition probabilities: (i) ``{\itshape growth probabilities}'' from any possible $\fancyR_{k-1}$, and (ii) a ``{\itshape  changepoint probability}'' from any possible $\fancyR_{k-1}$. In \cref{fig:bocpdDiagram}, growth transitions are represented by green diagonal lines whereas changepoint transitions are represented by red curves. At $\prescript{\Downarrow}{}{t_1}$, the algorithm evaluates only two hypotheses on $\fancyR_1$ in the form of joint probabilities: $P(\fancyR_1 = \zeta,\ {\bm o}(\prescript{\Downarrow}{}{t_1}))$ for $\zeta = 0 \text{ and } 1$.
To evaluate these hypotheses, the algorithm forward passes $P(\fancyR_0 = 0,\ {\bm o}(\prescript{\Downarrow}{}{t_0}) = \phi)$ from the previous time step and fuses it with the predictive probability. This step explains why the algorithm is often described as a ``{\itshape message-passing algorithm}''.
{\color{black} After finding the joint probabilities at $\prescript{\Downarrow}{}{t_1}$, the posterior distribution $P(\fancyR_1 = \zeta \ |\ {\bm o}(\prescript{\Downarrow}{}{t_1}))$ was determined using Bayes rule. Since the number of hypotheses on $\fancyR_k$ (possible values of $\zeta$) grows linearly over time steps, a new predictive model (corresponding to the new hypothesis) is concatenated to the set of underlying predictive models at the end of each Bayesian inference step. In subsequent inference steps, a predictive probability is computed for ${\bm o}(\prescript{\Downarrow}{}{t_k})$ given each underlying predictive model (or hypothesis) , unlike $\prescript{\Downarrow}{}{t_0}$ at which only the (non-)informative prior was available for prediction.}

\subsection*{Point Estimation of the Current Segment Run Length}

The aforementioned Bayesian estimation algorithm produced a posterior distribution over $\fancyR_k$ conditioned on the preprocessed joint kinematic timeseries so far observed at each time step $\prescript{\Downarrow}{}{t_k}$. However, $\fancyR_k$ is still a discrete random variable and each value it takes under this distribution corresponds to one run length hypothesis. Therefore, the use of point estimation is proposed to produce an estimate $\hat{\fancyR}_k$ that is close to the true value of $\fancyR_k$ in some probabilistic sense \cite{Dekking2005}. To this end, the {\itshape least mean squares} (LMS) estimator is employed (see \cref{eqn:eq26}) to produce $\hat{\fancyR}_k$ by minimising the {\itshape mean squared error} conditional on multiple continuous observations, ${\bm o}(\prescript{\Downarrow}{}{t_{1:k}})$.
{\color{black} The mean squared error is a effective criterion that trades-off bias and variance for an estimator.}
The output of the point estimation stage is a $(T+1)$ vector containing the LMS estimate of the run length at each time step.
\begin{equation}
\label{eqn:eq26}
\begin{split}
\hat{\fancyR}_k &= \arg \min_{\hat{\fancyR}_k} \mathbb{E}\left[(\fancyR_k - \hat{\fancyR}_k)^2\ |\ {\bm o}(\prescript{\Downarrow}{}{t_{1:k}})\right]\\
&= \mathbb{E}\left[\fancyR_k\ |\ {\bm o}(\prescript{\Downarrow}{}{t_{1:k}})\right]\\
&= \sum_{\fancyR_k} \fancyR_k\ P(\fancyR_k\ |\ {\bm o}(\prescript{\Downarrow}{}{t_{1:k}}))
\end{split}
\end{equation}

\subsection*{Reset Detection Logic for (In)activity Detection and Segmentation}

Encoded in $\hat{\fancyR}_k$ is the information on the segments and changepoints present in $\mathbfcal{O}(\prescript{\Downarrow}{}{\bm t})$. The preprocessed joint kinematic timeseries is now regarded as quasistatic segments of data which end every time $\hat{\fancyR}_k$ is reset upon body movement or activity. The detection of resets in $\hat{\fancyR}_k$ is necessary for the detection of changepoints which mark the start of active segments (or the end of inactivity).

\subsubsection*{Scale-invariant Reset Detection (Activity Detection)}

Reset detection is mainly about making a binary objective decision on whether the current segment of data samples is likely to continue growing or terminate. Importantly, the decision making needs to be invariant to the scale of $\hat{\fancyR}_k$, i.e. resets from $\hat{\fancyR}_k = \zeta_a$ and $\hat{\fancyR}_k = \zeta_b$ should both be treated equally by the algorithm even if $\zeta_b \gg \zeta_a$. Therefore, the application of threshold on the magnitude drop, $\hat{\fancyR}_k - \hat{\fancyR}_{k-1}$, would not meet the scale-invariant criterion. Alternatively, it is proposed to perform thresholding on the logarithmic scale of $\hat{\fancyR}_k$ since it effectively represents percent change or multiplicative factors. A reference drop of $\log_{10} 2 \approx 0.3$ on the logarithmic scale indicates halving of $\hat{\fancyR}_k$ in the linear scale, hence it constitutes a binary decision boundary for run length termination. Therefore, a reset is detected if the consecutive difference term, $\log_{10} \hat{\fancyR}_k - \log_{10} \hat{\fancyR}_{k-1}$, produced a drop larger than 0.3.

\subsubsection*{Postprocessing of Current Segment Run Length Point Estimates}

Prior to the detection of resets, it is recommended to incorporate a postprocessing algorithm to refine $\hat{\fancyR}_k$ for a better performance (refer to the Results Section). It was found that some gradual resets may occur, associated with few consecutive drops in $\hat{\fancyR}_k$. This resetting behaviour may be more challenging to be picked up by the reset detection logic. Therefore, a three-sample moving filter (defined in \cref{eqn:eq27}) was applied to $\hat{\fancyR}_k$ to merge these consecutive drops over one time step. The postprocessed estimate of $\hat{\fancyR}_k$ is denoted by $\hat{\fancyR}^p_k$:
\begin{equation}
\label{eqn:eq27}
\hat{\fancyR}^p_k =
\begin{cases}
\hat{\fancyR}_{k-1}, & \text{if $\hat{\fancyR}_{k-1} > \hat{\fancyR}_k > \hat{\fancyR}_{k+1}$.}\\
\hat{\fancyR}_k, & \text{otherwise.}
\end{cases}
\end{equation}

For clarity, if a two-consecutive-drop scenario is considered with $\hat{\fancyR}_{k-1} = 30$, $\hat{\fancyR}_k = 23$ and $\hat{\fancyR}_{k+1} = 0$, then the corresponding postprocessed LMS estimates will be $\hat{\fancyR}^p_{k-1} = 30$, $\hat{\fancyR}^p_k = 30$ and $\hat{\fancyR}^p_{k+1} = 0$. In this example, the moving filter produced a total drop of 30 between $\hat{\fancyR}^p_{k+1}$ and $\hat{\fancyR}^p_k$. Even though the changepoint would be detected one time step later than it actually was, the postprocessing algorithm prevents the larger risk of not detecting the changepoint event.

\subsubsection*{Elimination of Repetitive Resets During Activity}

During periods of activity, multiple resets may occur within a short time interval as the human participant makes few posture adjustments until they settle at the new sleep posture. Such repetitive resets do not allow $\hat{\fancyR}^p_k$ to proliferate in value, and hence they generally come after short run lengths. In this paper, the focus is more on sleep postures which are normally associated with longer run lengths (periods of inactivity). Consequently, the run length before each detected reset was thresholded to eliminate repetitive resets during periods of activity.
Any arbitrary reset at $\prescript{\Downarrow}{}{t_k}$ with $\hat{\fancyR}^p_{k-1} < 20$ was eliminated. The minimum run length of twenty samples was backed by histograms of $\hat{\fancyR}^p_k - \hat{\fancyR}^p_{k-1}$ from different participant datasets (example in \cref{fig:ADR_w_PP_Results}\hyperlink{fig:ADR_w_PP_Results_e}{e}).

\subsection*{Inactivity Segmentation}

After the filtration of repetitive resets, the remaining resets are those associated with sleep postures that persisted for sufficiently long run lengths. The duration of each posture is equal to the postprocessed point estimate of the current segment run length just before the detected reset. This was found to work better than taking the elapsed time between adjacent resets, simply because the Bayesian estimation of $\fancyR_k$ is a recursive and self-correcting algorithm. Therefore, the last posterior distribution, $P(\fancyR_k\ |\ {\bm o}(\prescript{\Downarrow}{}{t_{1:k}}))$, within a segment of inactivity is generally more reliable than previous posteriors.

\section*{Data availability}

The preprocessed joint kinematic timeseries belonging to participants can be accessed online on www.abcxyz.com.



\bibliography{arXiv_refs}



\section*{Acknowledgements (not compulsory)}

The authors would like to thank Daniel Potts and Shay Stanley for their assistance with the development of the wearable sensors used in the participant study.

\section*{Author contributions statement}

O. E. was responsible for the conceptualisation, research design, methodology and data acquisition. R. A. and A. H. contributed to the methodology and interpretation of data. A.H. instructed on the participant experimental protocol and research ethics. L. M. identified knowledge gaps from the literature, formulated the research problem and advised on the concept solution. F. C. and P. P. provided input on conceptualisation and research supervision. All authors edited and reviewed the manuscript, and approved the submitted version.



\section*{Competing interests}

O.E. is supported by the University of Liverpool Doctoral Network in AI for Future Digital Health. The authors declare no other competing interests.

\end{document}


\flushbottom
\makemytitle
%
%


{\noindent\bfseries\LARGE Supplementary Information}

\section*{Supplementary Methods - CG Pipeline}


Even though the axis-angle representation is intuitive, its four dimensions does not facilitate easy visualisation of joint orientation timeseries. Therefore, UMAP and analytical dimension reduction were used to reduce the dimensions to three. With fewer dimensions, the computational cost of subsequent Bayesian inference was lowered. This section is focused on the generation of synthetic orientation data essential for the UMAP-based dimension reduction.

Sensor-measured joint orientations are typically limited in size and does not cover the entire axis-angle orientation space due to anatomical constraints. For these reasons, forcing the UMAP to learn the data manifold of ${\bm x}$ from sensor measurements does not guarantee accurate manifold modelling and may suffer from mapping discontinuities.
Therefore, this paper proposes the generation of a synthetic axis-angle orientation dataset better suited for manifold learning by UMAP. Leveraging a CG pipeline, synthetic axes of rotation evenly distributed over the surface of a unit sphere were obtained. 
Then, an equidistant sequence of scalar angles of rotation was created. Finally, the synthetic orientation dataset was formed by exhausting all possible combinations between the axes and angles of rotation.

\begin{figure}[!t]
    \centering
    \begin{tikzpicture}
    \node[above left] (main_fig) at (0,0) {\includegraphics[width=0.7\textwidth]{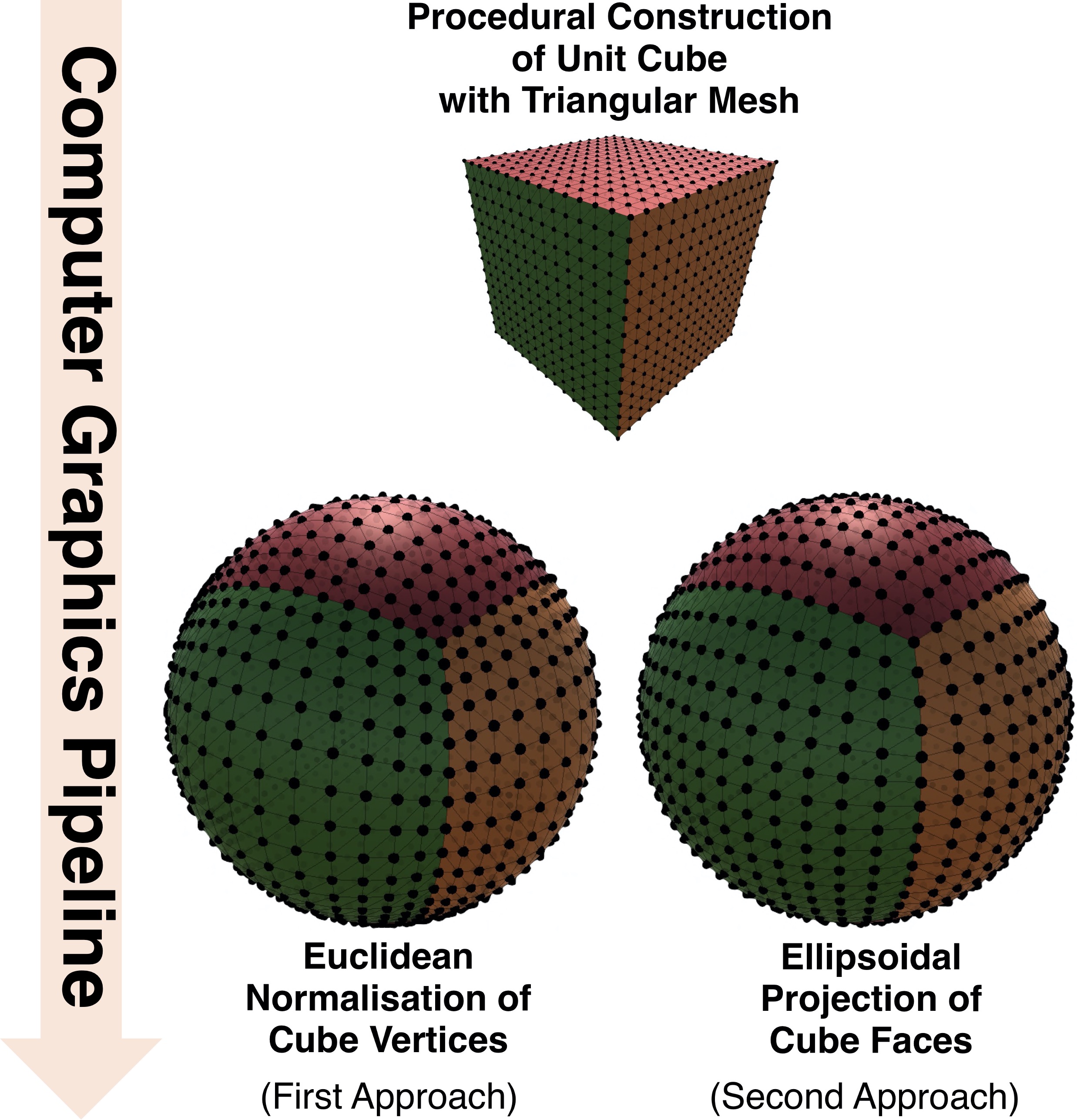}};
    
    \draw[ultra thick,rounded corners,black] (-0.475\linewidth,-0.04\textheight) rectangle (-0.475\linewidth+0.75cm,-0.04\textheight+0.75cm) node[pos=.5,align=center] {\hypertarget{fig:CG_pipeline_a}{\large a}};
    \draw[ultra thick,rounded corners,black] (-0.17\linewidth,-0.04\textheight) rectangle (-0.17\linewidth+0.75cm,-0.04\textheight+0.75cm) node[pos=.5,align=center] {\hypertarget{fig:CG_pipeline_b}{\large b}};
    \end{tikzpicture}
    
    \caption{\color{black}The computer graphics pipeline used for the generation of synthetic axes of rotation.}
    \label{fig:CG_pipeline}
\end{figure}

The CG pipeline aims at the procedural 3D modelling of a unit sphere with evenly-distributed vertices (axes of rotation).
As depicted in \cref{fig:CG_pipeline}, the pipeline generally consisted of two stages: (i) the construction of a unit cube with a structured triangular mesh, and (ii) the projection of the cube vertices onto the surface of a unit sphere.
All procedural 3D modelling was implemented in the C\# programming language and realised in Unity$^\copyright$ (Unity Technologies Inc., California, US).

The first stage of the CG pipeline involves procedural construction of the six faces making the unit cube. Each quadrilateral face is defined by a {\itshape surface normal vector}, $\vec{n}_{face}$, and a predefined {\itshape mesh resolution}, $\Upsilon$, such that $\Upsilon$ is the number of vertices along any side of the quadrilateral face. Given these definitions, the vertex grid of the face was constructed with $\Upsilon^2$ vertices, each assigned a unique identifier index.
The cube faces are all procedurally constructed using $\Upsilon=15$ and $\vec{n}_{face}$ being the six orthonormal directions of Unity's world coordinate system, which yields a total of 1,350 vertices defining the cube.

Next, the cube vertices need to be projected onto the surface of a unit sphere to obtain the synthetic axes of rotation. To this end, two approaches, based on {\itshape Euclidean normalisation} and {\itshape ellipsoidal projection} respectively, were employed and compared.

\subsection*{First Approach: Projection of Cube Vertices with Euclidean Normalisation}

Let ${\bm v}_c$ denote an arbitrary vertex on the unit cube, with individual coordinates $(x_c, y_c, z_c)$.
This approach projects ${\bm v}_c$ onto the surface of a unit sphere through the normalisation of ${\bm v}_c$ by its {\itshape Euclidean norm} as defined in \cref{eqn:eq3}.
\begin{equation}
\label{eqn:eq3}
\prescript{\text{euclid}}{}{{\bm v}_s} = \frac{1}{\sqrt{x_c^2 + y_c^2 + z_c^2}}
\cdot
\begin{bmatrix}
x_c &
y_c &
z_c 
\end{bmatrix}
\end{equation}

The result of this projection is displayed in \cref{fig:CG_pipeline}\hyperlink{fig:CG_pipeline_a}{a}, which clearly shows projected vertices clumped together at the corners of the faces than in the centres. Eventually, this projection would produce non-uniform samples over the data manifold in the orientation space, and could potentially cause unstable dimension reduction. Therefore, a second approach was developed to address this projection artefact.

\subsection*{Second Approach: Ellipsoidal Projection of Cube Faces}

\begin{figure}[!t]
    \centering
    \includegraphics[width=0.8\textwidth]{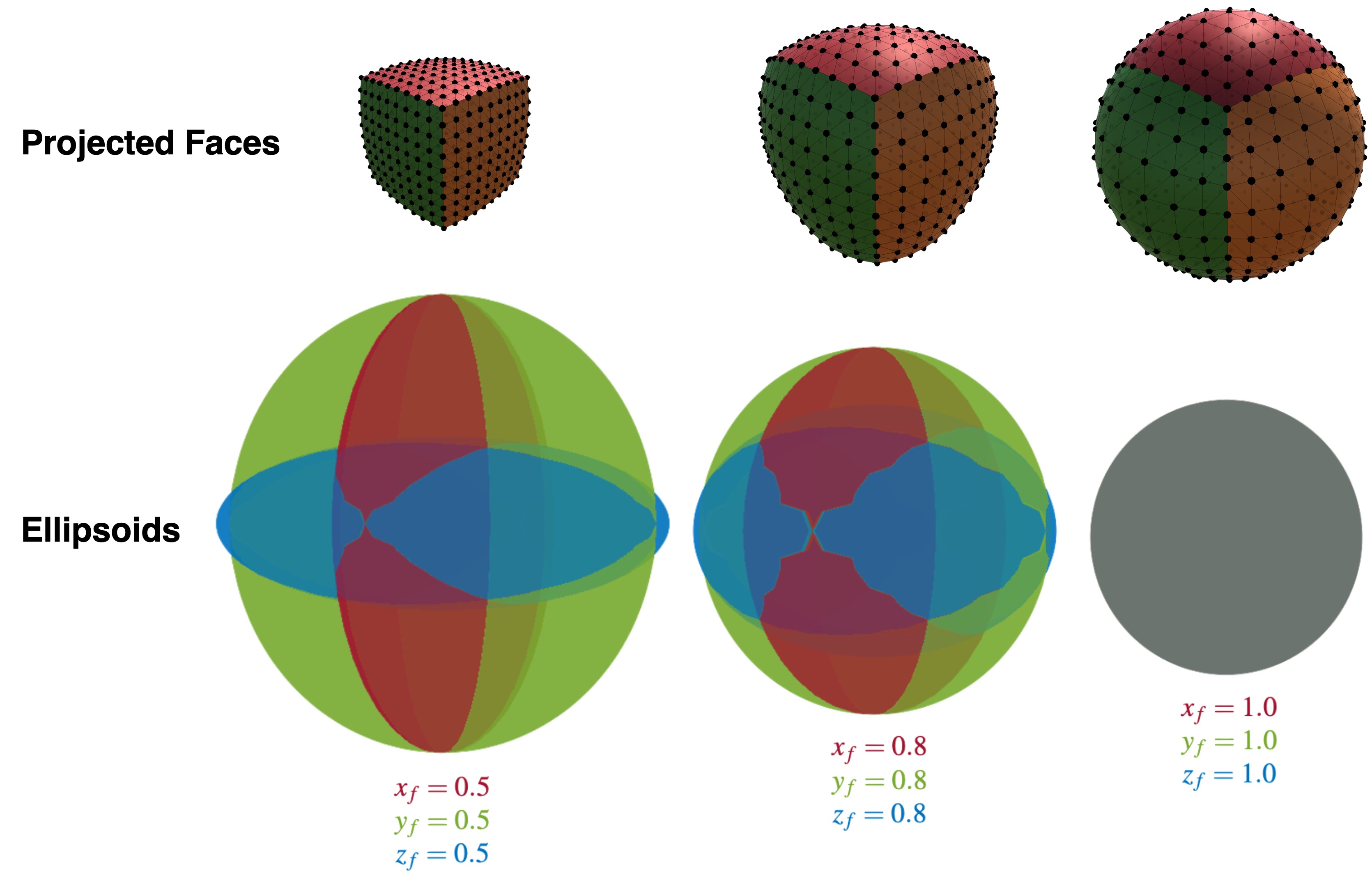}
    \caption{The effect of cube face displacement on the ellipsoidal projection.}
    \label{fig:ellipsoidal_projections}
\end{figure}

To obtain better distributed vertices, this approach projects constant-$x$, constant-$y$ and constant-$z$ square faces, which are the basic building geometries of a cube. The principal idea is to project each face onto an ellipsoid, and the stitched face ellipsoidal projections together make a unit sphere. Suppose a generic face vertex, ${\bm v}_f$, defined as
\begin{equation}
\label{eqn:eq4}
{\bm v}_f = \{(x_f,y_f,z_f)\ |\ x_f,y_f,z_f \in [-1,1]\}
\end{equation}
where either $x_f$, $y_f$ or $z_f$ is fixed at some constant value between -1 to 1, depending on the face orientation defined by its surface normal vector.
Mathematically, the ellipsoidal projection aims to map face vertices ${\bm v}_f$ to vertices ${\bm v}_e$ on a 3D ellipsoid such that the point of intersection with the coordinate axis remain unchanged, i.e. ${\bm v}_e = {\bm v}_f$ at the centre of the face, and the face vertices are more normalised the closer they are to the corners of the face.
For this requirement to hold true, the ellipsoid equation for some constant-$x_f$ face, as an example, must be
\begin{equation}
\label{eqn:eq5}
1 = \frac{x_e^2}{x_f^2} + \frac{y_e^2}{b_e^2} + \frac{z_e^2}{c_e^2}
\end{equation}
where $b_e$ and $c_e$ are constants, and $b_e = c_e$ due to symmetry about $x_e-y_e$ and $x_e-z_e$ planes.
As $x_f$ is varied from -1 to 1, the ellipsoid in \cref{eqn:eq5} make two intersection curves with the unit sphere, one curve for each of the upper and lower hemispheres. Any known point along these intersection curves can be used to determine $b_e$ and $c_e$. The upper intersection curve meets the $x_e-y_e$ plane at
\begin{equation}
\label{eqn:eq6}
{\bm v}_e =
\begin{bmatrix}
\ x_e & y_e & z_e\ 
\end{bmatrix}
=
\begin{bmatrix}
\ \frac{1}{\sqrt{2}}\ x_f & \frac{1}{\sqrt{2}}\sqrt{1-\frac{1}{2}\ x_f^2} & \frac{1}{\sqrt{2}}\sqrt{1-\frac{1}{2}\ x_f^2}\ 
\end{bmatrix}
\end{equation}
Substituting \cref{eqn:eq6} into \cref{eqn:eq5} yields $b_e = c_e = 2-x_f^2$. Hence, the ellipsoid for the constant-$x$ face becomes
\begin{equation}
\label{eqn:eq7}
1 = \frac{x_e^2}{x_f^2} + \frac{y_e^2}{2-x_f^2} + \frac{z_e^2}{2-x_f^2}
\end{equation}
By symmetry, the constant-$y$ and constant-$z$ faces are similarly mapped to the ellipsoids defined by \cref{eqn:eq8,eqn:eq9} respectively.
\begin{equation}
\label{eqn:eq8}
1 = \frac{x_e^2}{2-y_f^2} + \frac{y_e^2}{y_f^2} + \frac{z_e^2}{2-y_f^2}
\end{equation}
\begin{equation}
\label{eqn:eq9}
1 = \frac{x_e^2}{2-z_f^2} + \frac{y_e^2}{2-z_f^2} + \frac{z_e^2}{z_f^2}
\end{equation}

Then, the system of equations \cref{eqn:eq7,eqn:eq8,eqn:eq9} are simultaneously solved for $x_e$, $y_e$ and $z_e$ to obtain the general expression for the ellipsoidal projection, ${\bm v}_e$
\begin{equation}
\label{eqn:eq10}
{\bm v}_e
=
\begin{bmatrix}
x_e &
y_e &
z_e 
\end{bmatrix}
=
\begin{bmatrix}
\ x_f\sqrt{1 - \frac{1}{2}\ y_f^2 - \frac{1}{2}\ z_f^2 + \frac{1}{3}\ y_f^2z_f^2} &
y_f\sqrt{1 - \frac{1}{2}\ x_f^2 - \frac{1}{2}\ z_f^2 + \frac{1}{3}\ x_f^2z_f^2} &
z_f\sqrt{1 - \frac{1}{2}\ x_f^2 - \frac{1}{2}\ y_f^2 + \frac{1}{3}\ x_f^2y_f^2}\ 
\end{bmatrix}
\end{equation}

\cref{fig:ellipsoidal_projections} shows the ellipsoids for constant-$x$, constant-$y$ and constant-$z$ faces at different displacements from the origin of the Cartesian coordinate system. As $x_f$, $y_f$ and $z_f$ converge to 1.0, their corresponding ellipsoids converge in 3D shape to a unit sphere. This specific case of convergence applies to a cube with a side length of 1 unit, where the ellipsoidal projections of all six faces form a complete sphere when stitched together. Therefore, a better distributed vertex projection  $\prescript{\text{ellips}}{}{{\bm v}_s}$ (see \cref{fig:CG_pipeline}\hyperlink{fig:CG_pipeline_b}{b}) of the cube vertices onto the surface of a unit sphere was obtained using \cref{eqn:eq10}.

Each projected vertex on the surface of the sphere represents a candidate Cartesian axis of rotation. All 1,350 axes of rotation were written to a {\itshape comma-separated values} (CSV) file for subsequent import into MATLAB$^\copyright$ (The MathWorks, Massachusetts, US). Therein, an orientation dataset generator script concatenated each synthetic axis of rotation with each angle of rotation from the closed set $\{\frac{36}{36}\pi,\frac{35}{36}\pi,\frac{34}{36}\pi,...,\frac{1}{36}\pi\}$. The output synthetic axis-angle dataset contained a total of 48,600 orientations.

\section*{Supplementary Methods - Bayesian Estimation of Current Segment Run Length}

Based on the original paper \cite{Adams2007}, the general mathematical expression for the joint probability can be written as:
\begin{equation}
\label{eqn:eq14}
P(\fancyR_k,\ {\bm o}(\prescript{\Downarrow}{}{t_{1:k}})) =
\sum_{\fancyR_{k-1}}
\underbrace{P({\bm o}(\prescript{\Downarrow}{}{t_k})\ |\ \fancyR_{k},\ {\bm o}(\prescript{\Downarrow}{}{t_{1:k-1}}))}_{\text{\normalsize Predictive Term}}\ \ \ 
\underbrace{P(\fancyR_k\ |\ \fancyR_{k-1})}_{\text{\normalsize Changepoint Prior}}\ \ \ 
\underbrace{P(\fancyR_{k-1},\ {\bm o}(\prescript{\Downarrow}{}{t_{1:k-1}}))}_{\text{\normalsize Recursive Term}}
\end{equation}

{\bfseries The Predictive Term.} It is also referred to as the {\itshape posterior predictive term} because it predicts the next embedding given the previously observed embeddings and $\fancyR_k$. Thus, if a hypothesis states $\fancyR_k = \zeta$, only embeddings within the past $\zeta$ time steps should contribute to the predictive term:
\begin{equation}
\label{eqn:eq15}
P({\bm o}(\prescript{\Downarrow}{}{t_k})\ |\ \fancyR_k = \zeta,\ {\bm o}(\prescript{\Downarrow}{}{t_{1:k-1}})) =
P({\bm o}(\prescript{\Downarrow}{}{t_k})\ |\ \fancyR_k = \zeta,\ {\bm o}(\prescript{\Downarrow}{}{t_{k-\zeta:k-1}}))
\end{equation}
In this paper, it is assumed that $\mathbfcal{O}(\prescript{\Downarrow}{}{\bm t})$ is continuously distributed with unknown mean vector ${\bm \mu}_{\mathbfcal{O}} \in \mathbb{R}^3$ and precision matrix ${\bm \lambda}_{\mathbfcal{O}} \in \mathbb{R}^{3 \times 3}$. Additionally, a likelihood model, $P_l({\bm o}(\prescript{\Downarrow}{}{t_k})\ |\ {\bm \eta})$, is defined with model parameters ${\bm \eta} = \{{\bm \mu}_{\mathbfcal{O}},\ {\bm \lambda}_{\mathbfcal{O}}\}$.
To compute the predictive term, \cref{eqn:eq16} first finds the posterior distribution, $P({\bm \eta}\ |\ \fancyR_k = \zeta,\ {\bm o}(\prescript{\Downarrow}{}{t_{k-\zeta:k-1}}))$, then marginalises ${\bm \eta}$, conditional on $\fancyR_k$. Hence, the predictive term becomes the posterior predictive distribution:
\begin{equation}
\label{eqn:eq16}
P({\bm o}(\prescript{\Downarrow}{}{t_k})\ |\ \fancyR_k = \zeta,\ {\bm o}(\prescript{\Downarrow}{}{t_{k-\zeta:k-1}})) =
\int_{\bm \eta} P_l({\bm o}(\prescript{\Downarrow}{}{t_k})\ |\ {\bm \eta})\ \ \ 
P({\bm \eta}\ |\ \fancyR_k = \zeta,\ {\bm o}(\prescript{\Downarrow}{}{t_{k-\zeta:k-1}}))\ \ \ 
d{\bm \eta}
\end{equation}

The posterior distribution can be re-written as $P({\bm \eta}\ |\ \mathbfcal{D},\ {\bm \alpha})$, where $\mathbfcal{D} = {\bm o}(\prescript{\Downarrow}{}{t_{k-\zeta:k-1}})$ and ${\bm \alpha}$ denotes the parameters of the prior distribution, $P({\bm \eta}\ |\ {\bm \alpha})$.
Using Bayes' theorem, conditional probability and chain rule,  $P({\bm \eta}\ |\ \mathbfcal{D},\ {\bm \alpha})$ can be expanded as follows:
\begin{equation}
\label{eqn:eq17}
P({\bm \eta}\ |\ \mathbfcal{D},\ {\bm \alpha}) =
\frac{P_l(\mathbfcal{D}\ |\ {\bm \eta},\ {\bm \alpha})\ P({\bm \eta}\ |\ {\bm \alpha})}{P(\mathbfcal{D}\ |\ {\bm \alpha})}
\end{equation}

However, computing the integral in \cref{eqn:eq16} is intractable. Fortunately, {\itshape conjugate-exponential models} \cite{Murphy2007} provide an efficient solution to this problem. The concept of conjugacy allows for the derivation of a closed form expression as a function of ${\bm \alpha}$ without the need for integration. Hence, given a likelihood model $P_l(\mathbfcal{D}\ |\ {\bm \eta},\ {\bm \alpha})$, its conjugate prior model can be leveraged to simplify the posterior distribution to:
\begin{equation}
\label{eqn:eq18}
P({\bm \eta}\ |\ \mathbfcal{D},\ {\bm \alpha}) =
P({\bm \eta}\ |\ {\bm \alpha}^\prime)
\end{equation}
where both the prior and posterior distributions belong to the same distribution family with parameters ${\bm \alpha}$ and ${\bm \alpha}^\prime$ respectively.

Since both ${\bm \mu}_{\mathbfcal{O}}$ and ${\bm \lambda}_{\mathbfcal{O}}$ are inferred, the likelihood model follows a {\itshape multivariate Gaussian distribution}, for which the natural conjugate prior is the {\itshape Normal-Wishart distribution} with quadruple parameterisation ${\bm \alpha} = \{{\bm \mu}_\alpha,\ \kappa_\alpha,\ \nu_\alpha,\ {\bm \Sigma}_\alpha\}$,
\begin{equation}
\label{eqn:eq19}
\begin{split}
    P({\bm \eta}\ |\ {\bm \alpha}) &=
    \mathbfcal{NW}({\bm \eta}\ |\ {\bm \alpha}) \\
    &= \mathbfcal{N}({\bm \mu}_{\mathbfcal{O}}\ |\ {\bm \mu}_\alpha,\ (\kappa_\alpha\ {\bm \lambda}_{\mathbfcal{O}})^{-1})\ \ \ 
    \mathbfcal{W}({\bm \lambda}_{\mathbfcal{O}}\ |\ {\bm \Sigma}_\alpha,\ \nu_\alpha)
\end{split}
\end{equation}
where ${\bm \mu}_\alpha \in \mathbb{R}^3$ and $\kappa_\alpha  \in \mathbb{R}$ are the mean vector and the covariance scale coefficient, respectively, of the normal distribution, and $\nu_\alpha \in \mathbb{R}$ and ${\bm \Sigma}_\alpha \in \mathbb{R}^{3 \times 3}$ are the number of degrees of freedom and the scale matrix, respectively, of the Wishart distribution.

Since $P({\bm \eta}\ |\ {\bm \alpha})$ is a conjugate prior, then $P({\bm \eta}\ |\ \mathbfcal{D},\ {\bm \alpha})$ turns out to be a Normal-Wishart distribution too. Substituting $P({\bm \eta}\ |\ \mathbfcal{D},\ {\bm \alpha})$ into \cref{eqn:eq16}, it can be shown that the posterior predictive distribution follows a {\itshape Student's t-distribution} with $(\nu_\alpha^\prime - 2)$ degrees of freedom:
\begin{equation}
\label{eqn:eq20}
\begin{split}
P({\bm o}(\prescript{\Downarrow}{}{t_k})\ |\ \mathbfcal{D},\ {\bm \alpha}) &= 
\int P_l({\bm o}(\prescript{\Downarrow}{}{t_k})\ |\ {\bm \eta})\ \ \ 
P({\bm \eta}\ |\ \mathbfcal{D},\ {\bm \alpha})\ \ \ 
d{\bm \eta}\\
&= \mathbfcal{T}_{\nu_\alpha^\prime - 2}({\bm o}(\prescript{\Downarrow}{}{t_k})\ |\ {\bm \mu}_\alpha^\prime,\ {\bm \Sigma}_\alpha^\prime (\kappa_\alpha^\prime + 1) {\kappa_\alpha^\prime}^{-1} (\nu_\alpha^\prime - 2)^{-1})
\end{split}
\end{equation}
where ${\bm \mu}_\alpha^\prime$ and ${\bm \Sigma}_\alpha^\prime (\kappa_\alpha^\prime + 1) {\kappa_\alpha^\prime}^{-1} (\nu_\alpha^\prime - 2)^{-1}$ are the predicted ${\bm \mu}_{\mathbfcal{O}}$ and ${\bm \lambda}_{\mathbfcal{O}}^{-1}$, respectively, of the posterior predictive distribution.

{\bfseries Setting the prior on preprocessed joint kinematic timeseries.}
The set of parameters ${\bm \alpha}(k, \zeta)$ are updated at each time step $\prescript{\Downarrow}{}{t_k}$ and for each hypothesis $\fancyR_k = \zeta$.
The (non-)informative prior on ${\bm \eta}$ was defined by setting the initial parameters ${\bm \alpha}(0,0)$ of the prior model to the appropriate values.
In the case of the Normal-Wishart distribution, the non-informative prior, ${\bm \alpha}(0,0)$, was defined with ${\bm \mu}_\alpha(0,0) = \left[ 10^{-4}\ \ 10^{-4}\ \ 10^{-4} \right]$, $\kappa_\alpha(0,0) = 10^{-4}$, $\nu_\alpha(0,0) = 4$ and any matrix ${\bm \Sigma}_\alpha(0,0)$ provided that its determinant $\lvert{\bm \Sigma}_\alpha(0,0)\rvert = 0$. The non-informative prior is more suited for $\prescript{\text{UMAP}}{}{\mathbfcal{O}}$ due to the unconstrained embedding space of UMAP.

In contrast to UMAP, ADR has a constrained embedding space and known 3D embedding topology which facilitated the design of an informative prior. Visual inspection of the real preprocessed joint kinematics in the ADR embedding space informed the following settings: (1) ${\bm \mu}_{\mathbfcal{O}}$ has an expected value $E({\bm \mu}_{\mathbfcal{O}}) = 0$ and an initial guess of ${\bm I}_{3 \times 3}$ for its covariance matrix, and (2) ${\bm \lambda}_{\mathbfcal{O}}$ has an expected value $\mathbb{E}\left[{\bm \lambda}_{\mathbfcal{O}}\right] = \nu_\alpha\ {\bm \Sigma}_\alpha = 20\ {\bm I}_{3 \times 3}$.
Therefore, ${\bm \alpha}(0,0)$ was set such that ${\bm \mu}_\alpha(0,0) = \left[ 10^{-4}\ \ 10^{-4}\ \ 10^{-4} \right]$, $\kappa_\alpha(0,0) = \nicefrac{1}{20}$, $\nu_\alpha(0,0) = 4$ and ${\bm \Sigma}_\alpha(0,0) = 5\ {\bm I_{3 \times 3}}$.

{\bfseries Updating the prior model parameters.} At the end of each Bayesian inference step, the prior model parameters ${\bm \alpha}$ are updated to represent the new set of hypotheses on $\fancyR_k$.
At $\prescript{\Downarrow}{}{t_k}$ and given an arbitrary hypothesis $\fancyR_k = \zeta$, the prior model parameters denoted by ${\bm \alpha}(k, \zeta)$ are updated as follows:
\begin{equation}
\label{eqn:eq21}
\begin{split}
    {\bm \mu}_\alpha(k, \zeta) &=
    \frac{\kappa_\alpha(0,0) {\bm \mu}_\alpha(0,0) + \zeta\ \overline{\bm o}_{k-\zeta+1:k}}{\kappa_\alpha(0,0) + \zeta}\\
    \kappa_\alpha(k, \zeta) &=
    \kappa_\alpha(0,0) + \zeta\\
    \nu_\alpha(k, \zeta) &=
    \nu_\alpha(0,0) + \zeta\\
    {\bm \Sigma}_\alpha(k, \zeta) &= \!\begin{multlined}[t] {\bm \Sigma}_\alpha(0,0) + \frac{\kappa_\alpha(0,0)\ \zeta}{\kappa_\alpha(0,0) + \zeta} ({\bm \mu}_\alpha(0,0) - \overline{\bm o}_{k-\zeta+1:k})\ ({\bm \mu}_\alpha(0,0) - \overline{\bm o}_{k-\zeta+1:k})^T + \\
    \sum_{i=1}^\zeta ({\bm o}(\prescript{\Downarrow}{}{t_{k-i}}) - \overline{\bm o}_{k-\zeta+1:k})\ ({\bm o}(\prescript{\Downarrow}{}{t_{k-i}}) - \overline{\bm o}_{k-\zeta+1:k})^T
    \end{multlined}
\end{split}
\end{equation}
where $\overline{\bm o}_{k-\zeta+1:k}$ is the mean vector of the embeddings in ${\bm o}(\prescript{\Downarrow}{}{t_{k-\zeta+1:k}})$.

{\bfseries The Changepoint Prior.} This term allows for encoding the prior knowledge on changepoints present in $\mathbfcal{O}(\prescript{\Downarrow}{}{\bm t})$. The current segment run length $\fancyR_k$ is is a discrete, nonnegative random variable this is subject to two possible outcomes at any given $\prescript{\Downarrow}{}{t_k}$; either reset to zero upon a changepoint or an incremental growth by one. Thus, it was sensible to model $\fancyR_k$ with a geometric distribution of success probability, $p$, where success corresponds to the growth of $\fancyR_k$. Based on this modelling assumption, the changepoint prior term becomes
\begin{equation}
\label{eqn:eq22}
P(\fancyR_k\ |\ \fancyR_{k-1}) =
\begin{cases}
p, & \text{if $\fancyR_k = 0$.}\\
1-p, & \text{if $\fancyR_k = \fancyR_{k-1}+1$.}\\
0, & \text{otherwise.}
\end{cases}
\end{equation}

Using the changepoint prior in \cref{eqn:eq22}, the joint probability term in \cref{eqn:eq14} can be further simplified. For growth transitions, it is redundant to sum over all possible $\fancyR_{k-1}$ as the only valid run length is $\fancyR_k = \fancyR_{k-1} + 1$. Substituting \cref{eqn:eq18,eqn:eq22} into \cref{eqn:eq14}, the joint growth transition probability is defined as
\begin{equation}
\label{eqn:eq23}
P(\fancyR_k,\ {\bm o}(\prescript{\Downarrow}{}{t_{1:k}})) =
P({\bm o}(\prescript{\Downarrow}{}{t_k})\ |\ {\bm \alpha}(k-1, \fancyR_{k-1}))\ \ \ 
(1-p)\ \ \ 
P(\fancyR_{k-1},\ {\bm o}(\prescript{\Downarrow}{}{t_{1:k-1}})
\end{equation}

For the changepoint transitions, the run length can be reset from any valid $\fancyR_{k-1}$. Therefore, the summation over $\fancyR_{k-1}$ is necessary,
\begin{equation}
\label{eqn:eq24}
P(\fancyR_k,\ {\bm o}(\prescript{\Downarrow}{}{t_{1:k}})) =
\sum_{\fancyR_{k-1}} P({\bm o}(\prescript{\Downarrow}{}{t_k})\ |\ {\bm \alpha}(k-1, \fancyR_{k-1}))\ \ \ 
p\ \ \ 
P(\fancyR_{k-1},\ {\bm o}(\prescript{\Downarrow}{}{t_{1:k-1}})
\end{equation}

After finding all joint probabilities $P(\fancyR_k,\ {\bm o}(\prescript{\Downarrow}{}{t_{1:k}}))$ at $\prescript{\Downarrow}{}{t_k}$, the posterior distribution on $\fancyR_k$ was found using Bayes rule
\begin{equation}
\label{eqn:eq25}
P(\fancyR_k\ |\ {\bm o}(\prescript{\Downarrow}{}{t_{1:k}})) =
\frac{P(\fancyR_k,\ {\bm o}(\prescript{\Downarrow}{}{t_{1:k}}))}{\sum_{\fancyR_k} P(\fancyR_k,\ {\bm o}(\prescript{\Downarrow}{}{t_{1:k}}))}
\end{equation}
where $\sum_{\fancyR_k} P(\fancyR_k,\ {\bm o}(\prescript{\Downarrow}{}{t_{1:k}}))$ is the normalising evidence.

Once the posterior distribution over $\fancyR_k$ is determined $\forall k \in \left[0, T\right]$, this concludes the Bayesian inference of the current segment run length with $P(\fancyR_k\ |\ {\bm o}(\prescript{\Downarrow}{}{t_{1:k}}))$ being a $(T+1) \times (T+1)$ matrix indexed by $k$ and different $\fancyR_k=\zeta$. The upper triangular part of this matrix contains all-zero probabilities since $\fancyR_k \leq k$ at any $\prescript{\Downarrow}{}{t_k}$.



\bibliography{arXiv_refs}